\documentclass[showpacs,preprint,superscriptaddress,amsmath,amssymb]{revtex4-1}
\usepackage{graphicx}
\usepackage{dcolumn}
\usepackage{bm}
\usepackage{epsfig}
\usepackage{subfigure}
\usepackage{graphics}
\usepackage{amssymb}
\usepackage{amsthm}
\usepackage{amsmath}
\usepackage{array}
\usepackage{color}
\usepackage{booktabs}

\begin{document}

\preprint{APS/123-QED}

\title{Late-time description of immiscible Rayleigh-Taylor instability: A lattice Boltzmann study}

\author{Haowei Huang}
 \affiliation{Department of Physics, Hangzhou Dianzi University, Hangzhou 310018, China}

\author{Zhenhua Xia}
\affiliation{Department of Engineering Mechanics, Zhejiang University, Hangzhou 310027, China}

\author{Hong Liang}
\email[Corresponding author]{: xiazh@zju.edu.cn (Zhenhua Xia); lianghongstefanie@hdu.edu.cn (Hong Liang)}
\affiliation{Department of Physics, Hangzhou Dianzi University, Hangzhou 310018, China}%

\author{Yajing Zong}
\affiliation{Department of Physics, Hangzhou Dianzi University, Hangzhou 310018, China}

\author{Jiangrong Xu}
\affiliation{Department of Physics, Hangzhou Dianzi University, Hangzhou 310018, China}

\date{\today}

\begin{abstract}
In this paper, the late-time description of immiscible Rayleigh-Taylor instability (RTI) in a long duct is investigated over a comprehensive range of the Reynolds numbers ($1\leq Re \leq 10000$) and Atwood numbers $(0.05 \leq A \leq 0.7)$ based on the mesoscopic lattice Boltzmann method. We first reported that the instability with a moderately high Atwood number of 0.7 undergoes a sequence of distinguishing stages at a high Reynolds number, which are termed as the linear growth, saturated velocity growth, reacceleration and chaotic development stages. The spike and bubble at the secondary stage evolve with the constant velocities and their values agree well with the analytical solutions of the potential flow theory. The duration of the spike saturated velocity stage is very shorter than that of the bubble, implying the asymmetric developments between the spike and bubble fronts. Owing to the increasing strengths of the formed vortices, the movements of the spike and bubble are accelerated such that their velocities exceed than the asymptotic values and the evolution of the instability then enters into the reacceleration stage. Lastly, the curves for the spike and bubble velocities have some fluctuations at the chaotic stage and also a complex interfacial structure with large topological change is observed, while it still preserves the symmetry property with respect to the central axis. To determine the nature of the late-time growth, we also calculated the spike and bubble growth rates by using five popular statistical methods and two comparative techniques are recommended, resulting in the spike and bubble growth rates of about 0.13 and 0.022, respectively. When the Reynolds number is gradually reduced, some later stages such as the chaotic and reacceleration stages cannot be reached successively and the structure of the interface in the evolutional process becomes relatively smooth. The influence of the fluid Atwood number is also examined on the late-time RTI development at a large Reynolds number. It is found that the sustaining time for the saturated velocity stage decreases with the Atwood number. The spike late-time growth rate shows an overall increase with the Atwood number, while it has little influence on the bubble growth rate being approximately 0.0215.

\end{abstract}
\maketitle

\section{Introduction}
The Rayleigh-Taylor instability (RTI) is a fundamental interfacial instability between two fluids that occurs at a perturbed interface
when a heavy fluid is supported against gravity by a light fluid. This interesting instability phenomenon can be encountered in
cirrus cloud formation, accidental painting technique, and also has many relevant scientific and engineering applications such as
geophysics, inertial confinement fusion and astrophysics~\cite{Zhou1, Zhou2}. The pioneering work on the RTI was conducted
by Rayleigh and Taylor~\cite{Rayleigh}, who put forward the famous linear stability theory and showed that the RTI amplitude
grows exponentially with time. Later, Lewis~\cite{Lewis} experimentally validated the linear stability theory and found that
it was effective to describe the interface movement until its amplitude reaching to $0.4$ times the initial wavelength.
He also reported that the evolution of the RTI then entered into the nonlinear stage, where the bubble velocity was shown
to be proportional to the square root of the gravitational acceleration. The classical linear theory was derived for Euler flows,
and numerous subsequent researches have addressed the roles of density difference, viscosity, surface tension and compressibility
and the modified linear growth theories including these interwoven factors were deduced~\cite{Bellman, Mitchner, Plesset, Menikoff}.
Enlightened by these early works, the RTI subject has nourished the enthusiasm of mathematicians and scientists and interested readers
can refer to a good summary of early studies~\cite{Sharp} and the recent reviews~\cite{Zhou1, Zhou2, Boffetta, Livescu} on this topic.
	
Based on initial pattern of the perturbation, the RTI can be divided into the single-mode and multimode~\cite{Cabot, Adkins, Kobayashi} instabilities. The single-mode RTI in comparison to the multimode example is simple and ideal, but the study of single-mode phenomenon is still very important not only in its own interests to fluid mechanics, but also that the turbulent growth constant in the multimode instability has a direct relationship with the single-mode terminal velocity~\cite{Dimonte1}. Besides, the understanding of the single-mode phenomenon can provide good knowledge in the validation of the newly proposed numerical scheme, which thus is the focus of the current study. Following the linear growth stage,
the single-mode RTI would evolve with constant velocity, implying that the flow has been transformed into the nonlinear stage.
The nonlinear stage has been investigated by many analytical models~\cite{Layzer, Alon, Mikaelian, Zhang, Goncharov, Oron, Betti, Sohn, Banerjee}.
An important early contribution was attributed to Layzer~\cite{Layzer}, who theoretically analyzed the motion of the fluid-vacuum interface
in the gravitational field and proposed the first potential flow model for describing the RTI evolution with limiting Atwood number of 1.
Stating from Layzer's potential flow theory, several analytic formulas were subsequently presented for the nonlinear evolutions
of the bubble~\cite{Alon, Mikaelian} and spike~\cite{Zhang} amplitudes in the single-mode RTI. These models were only applicable
to the fluid-vacuum interfaces, and to remove this limitation, Goncharov~\cite{Goncharov} have generalized Layzer's theory to an improved potential
flow model for an arbitrary Atwood number, which predicts the saturated velocities of the inviscid spike $(u_s)$ and bubble $(u_b)$ as follows,
\begin{equation}
u_s=\sqrt{\frac{2Ag}{(1-A)Ck}},~~u_b=\sqrt{\frac{2Ag}{(1+A)Ck}},
\end{equation}
where $A$ is the Atwood number, $k$ is the perturbation
wave number, $g$ is the gravitational acceleration, $C$ is a constant given by 1 or 3 for the two-dimensional or
three-dimensional geometry. An alternative heuristic approach for describing the bubble and spike saturated velocities was
proposed by Oron {\it{et al.}}~\cite{Oron}, who constructed the buoyancy-drag model, interestingly yielding the identical
results to the potential flow approach of Goncharov~\cite{Goncharov}. Further, Betti and Sanz~\cite{Betti} have examined the
effect of the vorticity accumulation inside the bubble on the nonlinear evolution of ablative single-mode RTI, and the
resulting bubble saturated velocity including the vorticity effect was formulated. It is noticed that the viscosities of the fluids
are neglected in above potential flow models. Sohn~\cite{Sohn} and Banerjee~\cite{Banerjee} then analyzed the role of viscosity and
anticipated the increased skin-friction drag between viscous bubble and spike. To this end, the modified potential flow models
combining the viscosity effect and the vorticity~\cite{Banerjee} or the surface tension force~\cite{Sohn} were established for saturated velocity stage.

In addition to these theoretical studies, several researchers attempt to describe the nonlinear evolutional stages of the
single-mode RTI using the experimental and numerical approaches. Waddell {\it{et al.}}~\cite{Waddell} experimentally investigated
the two-dimensional single-mode RTI in low Atwood number fluid systems and reported that the initial growth rate agrees well with
linear stability theory. Additionally, they observed that average of the late-time spike and bubble velocities approaches to a constant.
Glimm {\it{et al.}}~\cite{Glimm} used the front-tracking method to carry out a long simulation of the two-dimensional single-mode RTI
with a moderate Atwood number. They found that the saturated velocity stage with a constant velocity breaks down at late time followed
by a new phenomena that the spike velocity was accelerated, indicating a reacceleration process after a plateau time. The reacceleration
stage was also confirmed in the numerical simulation of the single-mode RTI with low or moderate density
difference~\cite{Ramaprabhu} and the experiment in a low Atwood number system~\cite{Wilkison}. The reason
for the reacceleration behavior was ascribed to the formation of the vortices on the bubble-spike interfaces induced by the
Kelvin-Helmholtz (KH) instability. Ramaprabhu {\it{et al.}}~\cite{Ramaprabhu1} further conducted a longer simulation in order to
explore the late-time behaviours of the single-mode RTI, and first summarized the stability with low Atwood numbers
and large Reynolds numbers would undergo four different stages known as the linear growth, saturated velocity growth,
reacceleration, and chaotic mixing stages. During the chaotic mixing stage, the bubble and spike velocities experience an eventual
decrease lower than the saturated velocity of the potential flow model~\cite{Goncharov} and the symmetries within the bubble and spike images
were broken in their simulations. However, the later direct numerical simulations~\cite{Wei} indicated that the late-time bubble at high Reynolds number maintains a symmetric pattern and its growth exhibits a mean quadratic law. This late-time quadratic growth phenomenon was also observed in our mesoscopic numerical simulations of the single-mode RTI for low and moderate Atwood numbers~\cite{Liang1, Liang2}. Recently, Hu {\it{et al.}}~\cite{Hu1} numerically investigated the two-dimensional single-mode RTI with a low Atwood number and found that the bubble velocity for medium Reynolds numbers was decelerated and accelerated repeatedly after the reacceleration stage, they called this stage as the deceleration-acceleration stage. More recently, Bian {\it{et al.}}~\cite{Bian} conducted a detailed study on the growth of compressible single-mode RTI at different Reynolds numbers and Atwood numbers, and found that for a sufficiently high Reynolds number, the bubble late-time acceleration is persistent and does not vanish, which is consistent with the direct numerical results of incompressible RTI~\cite{Wei}.
	
As reviewed above, there have been some numerical works~\cite{Ramaprabhu1, Wei, Liang1, Liang2, Hu1, Bian} to
investigate the late-time single-mode RTI, which provide a good understanding of this complex interfacial instability phenomenon.
However, most of these studies~\cite{Wei, Liang1, Liang2, Hu1} only focus on the low Atwood number fluid systems, and the instability durations in all reported simulations are relatively short (the aspect ratio of the two-dimensional computational domain does not exceed than 8). Actually, it needs a longer vertical domain for the flow to reach the late-time stage for a high Atwood number. In addition, the quantitative description on the late-time growth of the single-mode RTI is still very lacking. To fill this gap, we invoked an improved mesoscopic lattice Boltzmann algorithm~\cite{Liang3} to systematically explore the late-time dynamics of the single-mode RTI in an elongated duct with an aspect ratio of 15, and the influences of wide Reynolds numbers and Atwood numbers are investigated in detail. The rest of this article is organized as follows. In Sec.~\ref{sec:method}, we will give a brief introduction of the used numerical methods. Sec~\ref{sec:Results} describes the long-time evolution of single-mode instability with different Reynolds numbers and Atwood numbers. Finally, we conduct a summary in Sec.~\ref{sec:summary}.

\section{Mathematical method}~\label{sec:method}
The phase-field model~\cite{Jacqmin, Shen} has recently received great attention in simulating multiphase flows due
to its firm physics for describing interfacial dynamics. In this model, an order parameter is introduced to identify different phases,
and the interface is assumed as the transition region of finite thickness where the physical variables are smoothly distributed. This feature makes it become advantageous for tackling interfacial flows involving large topological change. Considering the space $\Omega$ occupied by a multiphase system, the total free-energy functional of a multiphase system can be expressed by~\cite{Jacqmin},
\begin{equation}
E=\int_\Omega [F(\phi)+{\frac{k}{2}|\nabla\phi|^2}] d\Omega,
\end{equation}
where $\phi$ is an order parameter, $F(\phi)$ is the bulk free-energy density taking the form of the double-well
potential $F(\phi)=\beta\phi^2(1-\phi)^2$, ${\frac{k}{2}|\nabla\phi|^2}$ accounts for the excess free energy in the interfacial
region, the parameters $k$, $\beta$ depend on the surface tension $\sigma$ and the interface
width $D$ by $D=\sqrt{8k/\beta}$, $\sigma=\sqrt{2k\beta}/6$. The Cahn-Hilliard model admits the energy dissipation law
and is viewed as a gradient flow with respect to the free-energy functional~\cite{Shen}. In this case, the chemical potential can
be defined as the variation of total free-energy functional (2),
\begin{equation}
\mu = 4\beta \phi(\phi - 1)(\phi - \frac{1}{2})-k{\nabla ^2}\phi,
\end{equation}
and then the classical Cahn-Hilliard equation for governing the order parameter can be expressed by
\begin{equation}
{{\partial \phi } \over {\partial t}} + \nabla  \cdot (\phi {\bf{u}})
= \nabla  \cdot {M}(\nabla {\mu} ),
\end{equation}
where ${\bf{u}}$ is the fluid velocity, $M$ is the mobility. To simulate interfacial flows, the Cahn-Hilliard equation
should be coupled with the incompressible Navier-Stokes equations including the surface tension (${\mathbf{F}}_s$) and
external forces ($\mathbf{G}$),
\begin{subequations}
\begin{equation}
\nabla  \cdot {\bf{u}} = 0,
\end{equation}
\begin{equation}
\rho ({{\partial {\bf{u}}} \over{\partial t}} + {\bf{u}}\cdot\nabla {\bf{u}} )=  - \nabla p +\nabla  \cdot \left[ {\nu\rho(\nabla {\bf{u}} + \nabla{{\bf{u}}^T})} \right] + {{\bf{F}}_s} + {\bf{G}},
\end{equation}
\end{subequations}
where $\rho$ is the fluid density, $p$ is the hydrodynamic pressure, $\nu$ is the kinematic viscosity. To achieve a low spurious velocity,
we choose the potential form of the surface tension force $\mathbf{F}_s=\mu\nabla{\phi}$~\cite{Jacqmin}.
	
To solving the coupled system of the Cahn-Hilliard and Navier-Stokes equations, some mesoscopic lattice Boltzmann algorithms~\cite{Liang1, Zu, Liang4} have been reported, among which an improved one~\cite{Liang1} is utilized to investigate the variable-density RTI owing to its great potential in solving complex interfacial instability~\cite{Liang5}. The mesoscopic algorithm takes advantages of double distribution functions and their evolutional equations equipped with an advanced multiple-relaxation-time collision operator can be written as~\cite{Guo}
\begin{subequations}
\begin{equation}
{f}_i({\bf{x}} +{{\bf{c}}_i}{\delta _t},t + {\delta _t}) -{f}_i({\bf{x}},t) = -({\bf{M}}^{-1}{\bf{S}}_{f}{\bf{M}})_{ij}[{{f}_j({\bf{x}},t) -f_j^{eq}({\bf{x}},t)}]+ {\delta _t}[\mathbf{M}^{-1}(\mathbf{I}-\frac{\mathbf{S}_f}{2})\mathbf{M}]_{ij}{F_j({\bf{x}},t)},
\end{equation}
\begin{equation}
{g}_i({\bf{x}} +{{\bf{c}}_i}{\delta _t},t + {\delta _t}) -{g}_i({\bf{x}},t) = -({\bf{M}}^{-1}{\bf{S}}_{g}{\bf{M}})_{ij}[{{g}_j({\bf{x}},t) -g_j^{eq}({\bf{x}},t)}]+ {\delta_t}[\mathbf{M}^{-1}(\mathbf{I}-\frac{\mathbf{S}_g}{2})\mathbf{M}]_{ij}{G_j({\bf{x}},t)},
\end{equation}
\end{subequations}
where ${f}_i({\bf{x}},t)$ and ${g}_i({\bf{x}},t)$ are the particle distribution functions at position $\mathbf{x}$
and time $t$, $f_i^{eq}({\bf{x}},t)$ and $g_i^{eq}({\bf{x}},t)$ are the equilibrium distribution functions, ${\bf{c}}_i$ is
the discrete velocity, ${\bf{M}}$ is collision matrix, ${\bf{S}}_f$ and ${\bf{S}}_g$ are the diagonal relaxation matrices,
$F_i({\bf{x}},t)$ and $G_i({\bf{x}},t)$ are the distribution functions of source term and forcing term, respectively. To be consistent with
the target equations, the equilibrium distributions are designed as~\cite{Liang1}
\begin{equation}
f_i^{eq}=\left\{
\begin{array}{ll}
\phi+({\omega_i} - 1)\eta\mu,              & \textrm{ $i=0$},    \\
{\omega_i}\eta\mu+{\omega_i}\phi{{{\bf{c}}_i \cdot {\bf{u}}}\over {c_s^2}},  & \textrm{ $i\neq0$} \\
\end{array}
\right.
\end{equation}
and
\begin{equation}
g_i^{eq}=\left\{
\begin{array}{ll}
{p \over {c_s^2}}({\omega _i} - 1) + \rho{s_i}({\bf{u}}),              & \textrm{ $i=0$},    \\
{p \over {c_s^2}}{\omega _i} + \rho{s_i}({\bf{u}}),                    & \textrm{ $i\neq0$} \\
\end{array}
\right.
\end{equation}
with
\begin{equation}
{s_i}({\bf{u}}) = {\omega_i}\left[ {{{{{\bf{c}}_i} \cdot {\bf{u}}}\over {c_s^2}} + {{{{({{\bf{c}}_i} \cdot {\bf{u}})}^2}} \over{2c_s^4}} - {{{\bf{u}} \cdot {\bf{u}}} \over {2c_s^2}}} \right],
\end{equation}
where $\eta$ is a free parameter that controls the mobility, $c_s$ is the sound speed, and $\omega_i$ is the weighting coefficient.
For two-dimensional flows, we apply the popular $D2Q9$ discrete-velocity model~\cite{YWei1, Chai, YWei2}
to the lattice Boltzmann algorithm, where the weighting coefficient $\omega_i$ is given by $\omega_0=4/9$, $\omega_{1-4}=1/9$ and $\omega_{5-8}=1/36$,
and the discrete velocity ${\bf{c}}_i$ is defined by
\begin{equation}
\mathbf{c}_{i}=\left\{
\begin{array}{ll}
(0,0)c,                                                         & \textrm{ $i=0$},   \\
(\cos [(i-1)\pi /2],\sin [(i-1)\pi /2])c,                       & \textrm{ $i=1-4$}, \\
\sqrt{2}(\cos [(i-5)\pi /2+\pi /4],\sin [(i-5)\pi /2+\pi /4])c, & \textrm{ $i=5-8$},
\end{array}
\right.
\end{equation}
where $c=\delta _x/\delta _t$, $\delta _x$ and $\delta _t$ are the lattice length and time step respectively,
and $c_s=c/\sqrt{3}$. In the multiphase lattice Boltzmann method, $\delta_x$ and $\delta_t$ are conventionally set to
be length and time units, i.e., $\delta_x=\delta_t=1.0$. The corresponding collision matrix $\bf{M}$ for the $D2Q9$ model
is given as~\cite{Lallemand}
\begin{displaymath}
\bf{M}= \left(
\begin{array}{rrrrrrrrr}
  1 & 1 & 1 & 1 & 1 & 1 & 1 & 1 & 1 \\
  -4 & -1 & -1 & -1 & -1 & 2 & 2 & 2 & 2 \\
  4 & -2 & -2 & -2 & -2 & 1 & 1 & 1 & 1 \\
  0 & 1 & 0 & -1 & 0 & 1 & -1 & -1 & 1 \\
  0 & -2 & 0 & 2 & 0 & 1 & -1 & -1 & 1 \\
  0 & 0 & 1 & 0 & -1 & 1 & 1 & -1 & -1 \\
  0 & 0 & -2 & 0 & 2 & 1 & 1 & -1 & -1 \\
  0 & 1 & -1 & 1 & -1 & 0 & 0 & 0 & 0 \\
  0 & 0 & 0 & 0 & 0 & 1 & -1 & 1 & -1
\end{array}\right),
\end{displaymath}
which is used to transform the particle distribution functions into their corresponding moments. With some algebraic manipulations, the equilibrium distribution functions in the moment space can be given by
\begin{subequations}
\begin{equation}
\mathbf{m}^{eq}_f=(\phi, -4\phi+2\eta\mu, 4\phi-3\eta\mu, \phi u_x, -\phi u_x, \phi u_y, -\phi u_y, 0, 0)
\end{equation}
\begin{equation}
\mathbf{m}^{eq}_g=(0, 6p+3\rho|\mathbf{u}|^2, -9p-3\rho|\mathbf{u}|^2, \rho u_x, -\rho u_x, \rho u_y, -\rho u_y, \rho(u_x^2-u_y^2), \rho u_x u_y)
\end{equation}
\end{subequations}
where $u_x$, $u_y$ are the horizontal and vertical components of velocity. The relaxation matrices of ${\bf{S}}_f$ and ${\bf{S}}_g$ in
Eqs. (6a) and (6b) are defined by
\begin{gather}
{\mathbf{S}}_f=diag(s_0^f,s_1^f,s_2^f,s_3^f,s_4^f,s_5^f,s_6^f,s_7^f,s_8^f), \\
{\mathbf{S}}_g=diag(s_0^g,s_1^g,s_2^g,s_3^g,s_4^g,s_5^g,s_6^g,s_7^g,s_8^g),
\end{gather}
where $0<s_i^f<2, 0<s_i^g<2$. Another indispensable event in the lattice Boltzmann approach is to design a suitable forcing
distribution function since the discrete lattice effect could be induced when the source or forcing term is incorporated.
Previously, Liang {\it{et al.}}~\cite{Liang1} considered the discrete lattice effect and added the time derivative term into
the source distribution function
\begin{equation}
F_i=\frac{\omega_i\mathbf{c}_{i}\cdot\partial_t(\phi\mathbf{u})}{c_s^2}
\end{equation}
and also in order to recover the incompressible Navier-Stokes equations correctly, the forcing distribution function $G_i$ in Eq. (6b)
is defined by~\cite{Liang1}
\begin{equation}
{{G}_i}=\frac{(\mathbf{c}_i-\mathbf{u})}{c_s^2}\cdot[s_i(\mathbf{u})\nabla(\rho{c_s^2})+(\mathbf{F}_s+\mathbf{F}_a+\mathbf{G})(s_i(\mathbf{u})+\omega_i)],	\end{equation}
where $\mathbf{F}_a$ is an additional interfacial force given by $\mathbf{F}_a=\frac{d\rho}{d \phi}M\nabla^2\mu \mathbf{u}$.
The macroscopic quantities $\phi$, $\bf{u}$ and $p$ can be obtained by calculating zeroth- or first-order moments of the particle
distribution functions,
\begin{subequations}
\begin{equation}
\phi = \sum\limits_i {{f_i}},
\end{equation}
\begin{equation}
\rho{\bf{u}} ={{\sum\limits_i {{{\bf{c}}_i}{g_i}} + \frac{{\delta_t}}{2}({{\bf{F}_s}+\mathbf{F}_a+\mathbf{G}})}},
\end{equation}
\begin{equation}
p = {{c_s^2} \over {(1 - {\omega_0})}}[ {\sum\limits_{i \ne 0} {{g_i}}  + {{{\delta _t}} \over 2}{\bf{u}} \cdot \nabla \rho + \rho {s_0}(\textbf{u})}],
\end{equation}
\end{subequations}
and the macroscopic density $\rho$ is determined by a linear differential function with respect to the order parameter,
\begin{equation}
\rho=\phi(\rho_h-\rho_l)+\rho_l,
\end{equation}
where $\rho_h$ and $\rho_l$ are the densities of the heavy and light fluids. Applying the Chapman-Enskog procedure,
the used lattice Boltzmann method can recover the exact Cahn-Hilliard and Navier-Stokes coupled equations with the mobility $M$ and the kinematic viscosity $\nu$ given by
\begin{subequations}
\begin{equation}
M=\eta c_s^2(\tau_f-0.5)\delta_t.
\end{equation}
\begin{equation}
\nu=c_s^2(\tau_g-0.5)\delta_t,
\end{equation}
\end{subequations}
where $1/\tau_f=s_3^f=s_5^f$, and $1/\tau_g=s_7^g=s_8^g$. In our simulations, the relation $s_0^g=s_3^g=s_5^g=1.0$ is used,
$s_7^g$ and $s_8^g$ are adjusted according to the value of the Reynolds number. The remaining relaxation parameters
in Eq. (13) are chosen as $s_1^g=s_2^g=1.0$ and $s_4^g=s_6^g=1.7$, while all the relaxation factors $s_i^f$ in Eq. (12)
are given as 1.0. Besides, in the practical applications, the temporal and spacial derivatives need to be numerically calculated, here we adopt the explicit Euler scheme to compute the time derivative, i.e., $\partial_t(\phi \mathbf{u})=[\phi(t)\mathbf{u}(t)-\phi(t-\delta_t)\mathbf{u}(t-\delta_t)]/\delta_t$, and the spacial gradient terms
are discretized by a second-order isotropic central schemes,
\begin{equation}
\nabla \chi({\bf{x}})=\sum\limits_{i \ne 0}{\frac{\omega_i\textbf{c}_i\chi({\bf{x}} +{{\bf{c}}_i}{\delta_t})}{c_s^2 \delta_t}}
\end{equation}
\begin{equation}
\nabla^2\chi({\bf{x}})=\sum\limits_{i \ne 0}{\frac{2\omega_i[\chi({\bf{x}} +{{\bf{c}}_i}{\delta_t})-\chi({\bf{x}})]}{c_s^2 \delta_t^2}},
\end{equation}
where $\chi$ represents an arbitrary variable that the derivative operator has been applied.

\section{Numerical Results and Discussions}~\label{sec:Results}

To reach the late-time stage for a high Atwood number, we consider the evolution of the single-mode instability in a sufficiently long
rectangular pipe divided by an even mesh of $L\times W=8160\times 544$, which has already given the grid-independent results. Initially, a small cosine perturbation is seeded on the central section of the rectangle and then the initial phase interface is located at
\begin{equation}
h(x,y)=0.5L+0.05W\cos(k\pi),
\end{equation}
where the wave number $k$ is defined by $k=2\pi/W$. To match the pattern of the interface, the initial distribution
for the order parameter is given by
\begin{equation}
\phi(x,y)=0.5+0.5\tanh \frac{2[y-h(x,y)]}{D},
\end{equation}
where the interface thickness $D$ is set to be four lattice units, and the surface tension $\sigma$ is fixed as $5\times10^{-5}$.
As is well known, the behaviour of the single-mode instability can be characterized by two important dimensionless numbers,
i.e., the Reynolds number ($Re$) and the Atwood number($A$), which are defined as follows~\cite{Hex},
\begin{equation}
Re=\frac{W\sqrt{gW}}{\nu},~~A=\frac{\rho_h-\rho_l}{\rho_h+\rho_l},
\end{equation}
where $\nu$ is the kinematic viscosity. To incorporate the gravitational effect, the following buoyancy force $\mathbf{G}$
in the vertical direction is applied to
binary fluids,
\begin{equation}
\mathbf{G}=[0, -(\rho-\frac{\rho_h+\rho_l}{2})g].
\end{equation}
In the simulations, the density of the heavy fluid ($\rho_h$) is set as 1.0 and the density for the light fluid ($\rho_l$)
can be determined by a given Atwood number. The gravitational acceleration is chosen to satisfy the relation $\sqrt{g W}=0.04$
and the Peclet number $(Pe)$ determining the value of the mobility is fixed as 50 based on the definition $Pe=\sqrt{g W} D/M\beta$~\cite{Zu}.
The periodic boundary conditions are applied in the horizontal direction, while the halfway bounce-back scheme is used on the upper
and lower boundaries. In this study, the characteristic length is chosen as the the wavelength of the perturbation $W$, $\sqrt{gW}$
is regarded as the characteristic velocity, and then the characteristic time is defined by $\sqrt{W/g}$. The present definitions of
the characteristic velocity and time are different from the ones $\sqrt{\frac{AgW}{1+A}}$ and $\sqrt{\frac{(1+A)W}{Ag}}$ used in previous
literatures~\cite{Wei, Hu1}, which obviously depend on the studied density difference of two fluids and would be unreasonable from
the perspective of physics. In addition, we would like to stress that the relevant statistics presented below have been normalized
by these characteristic values.

\subsection{Effect of the Reynolds number}

\begin{figure}
\subfigure{\includegraphics[width=2.9in,height=2.8in] {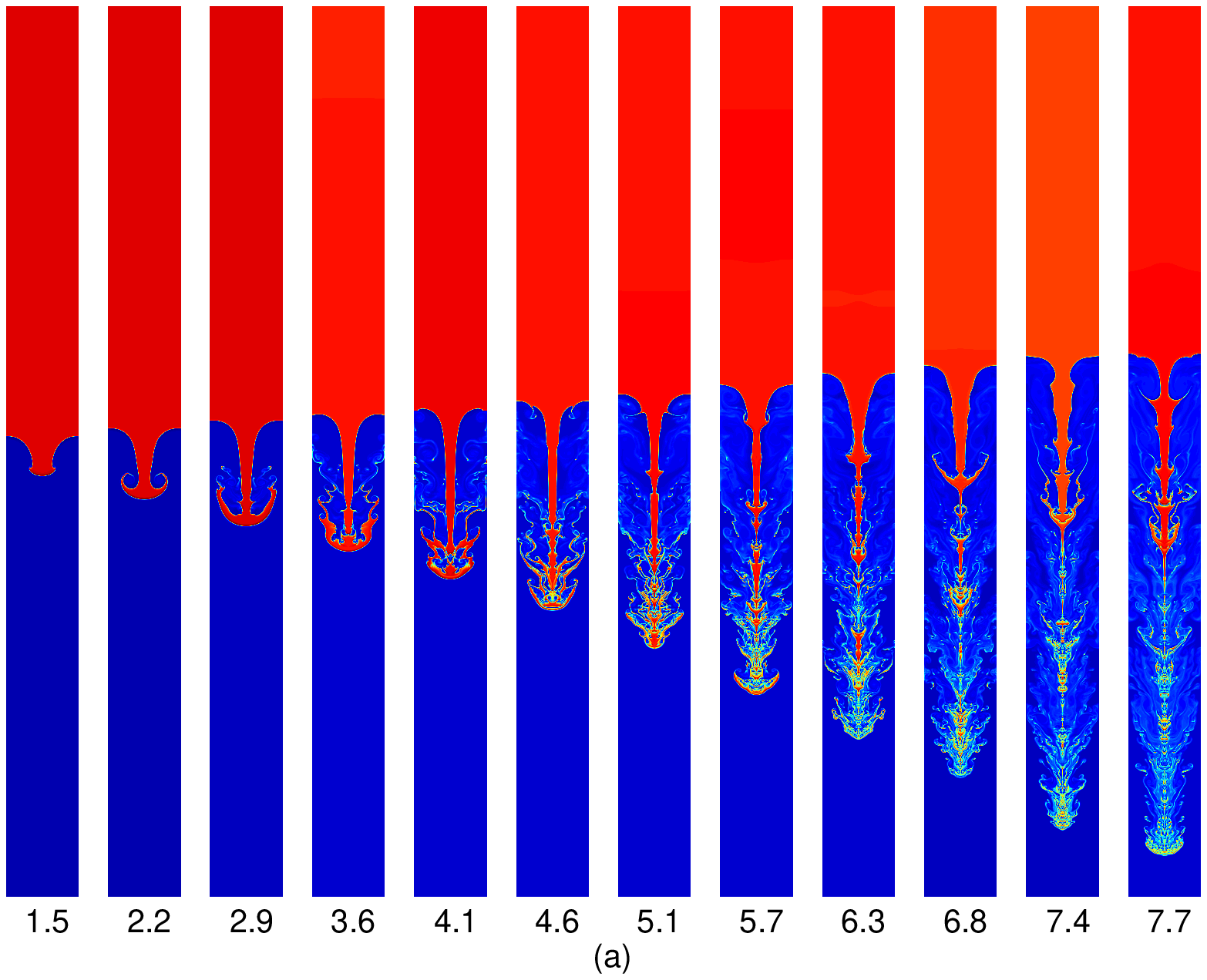} }\hspace{0.7cm}
\subfigure{\includegraphics[width=2.9in,height=2.8in] {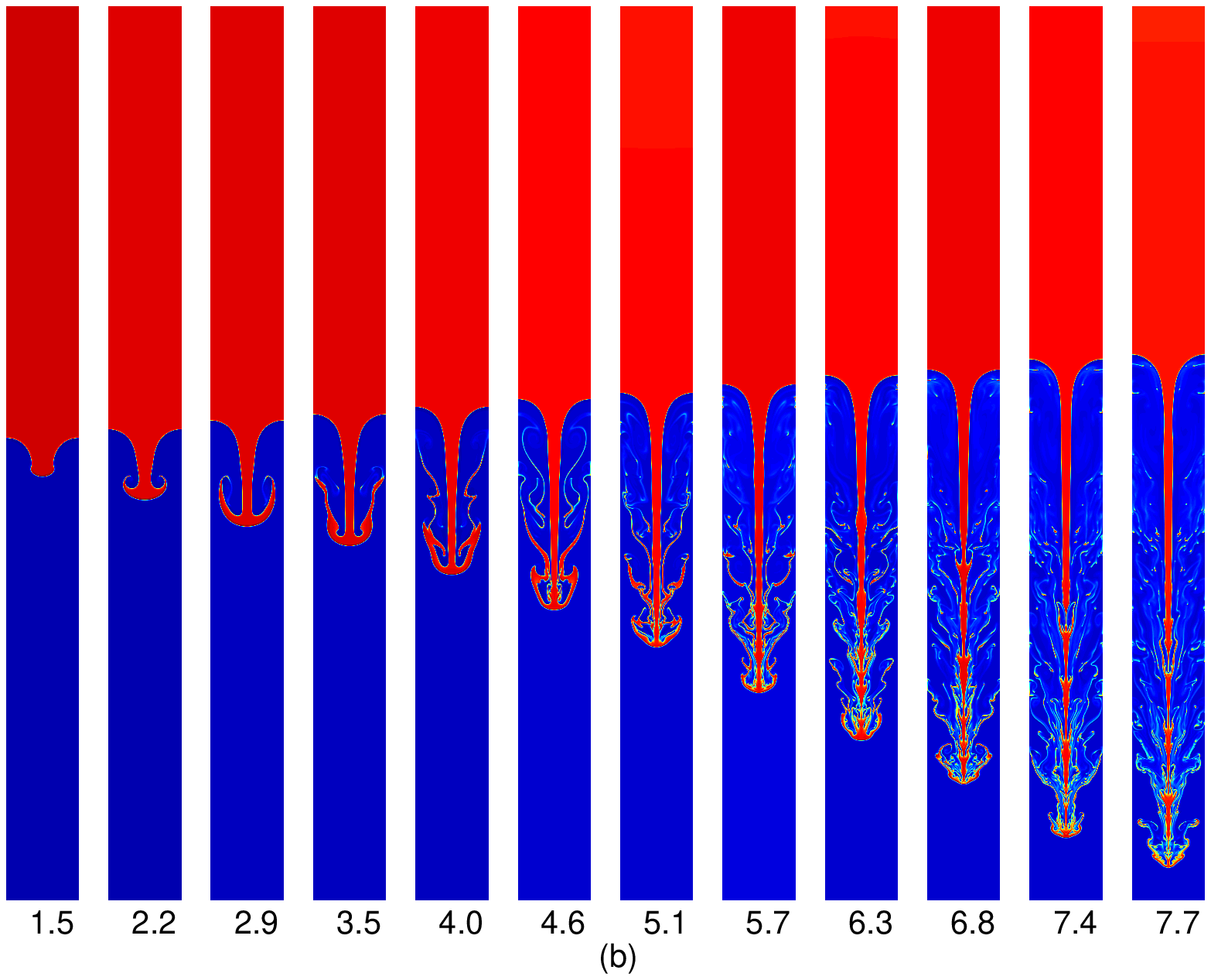}}\vskip 16pt
\subfigure{\includegraphics[width=2.9in,height=2.8in] {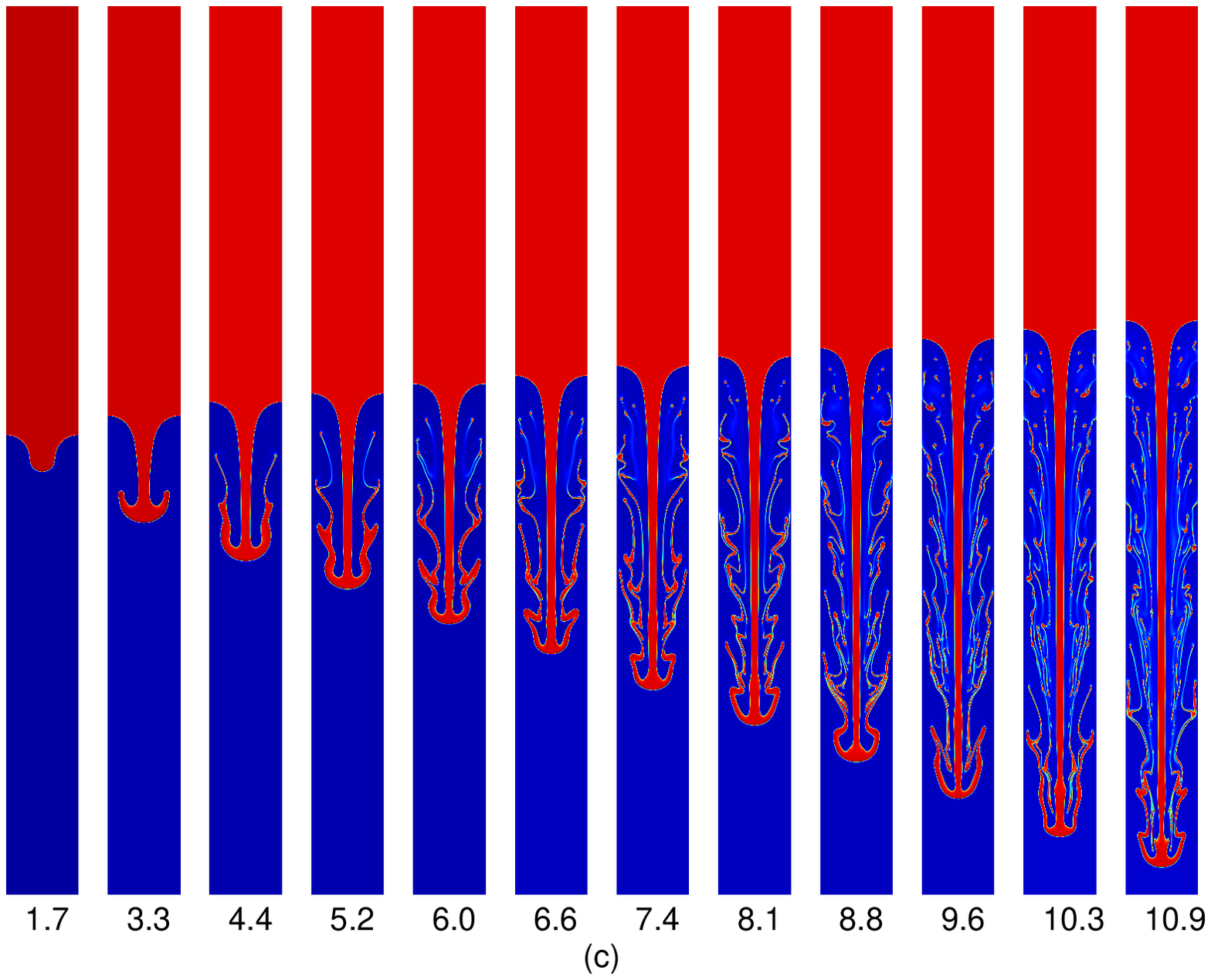} }\hspace{0.7cm}
\subfigure{\includegraphics[width=2.9in,height=2.8in] {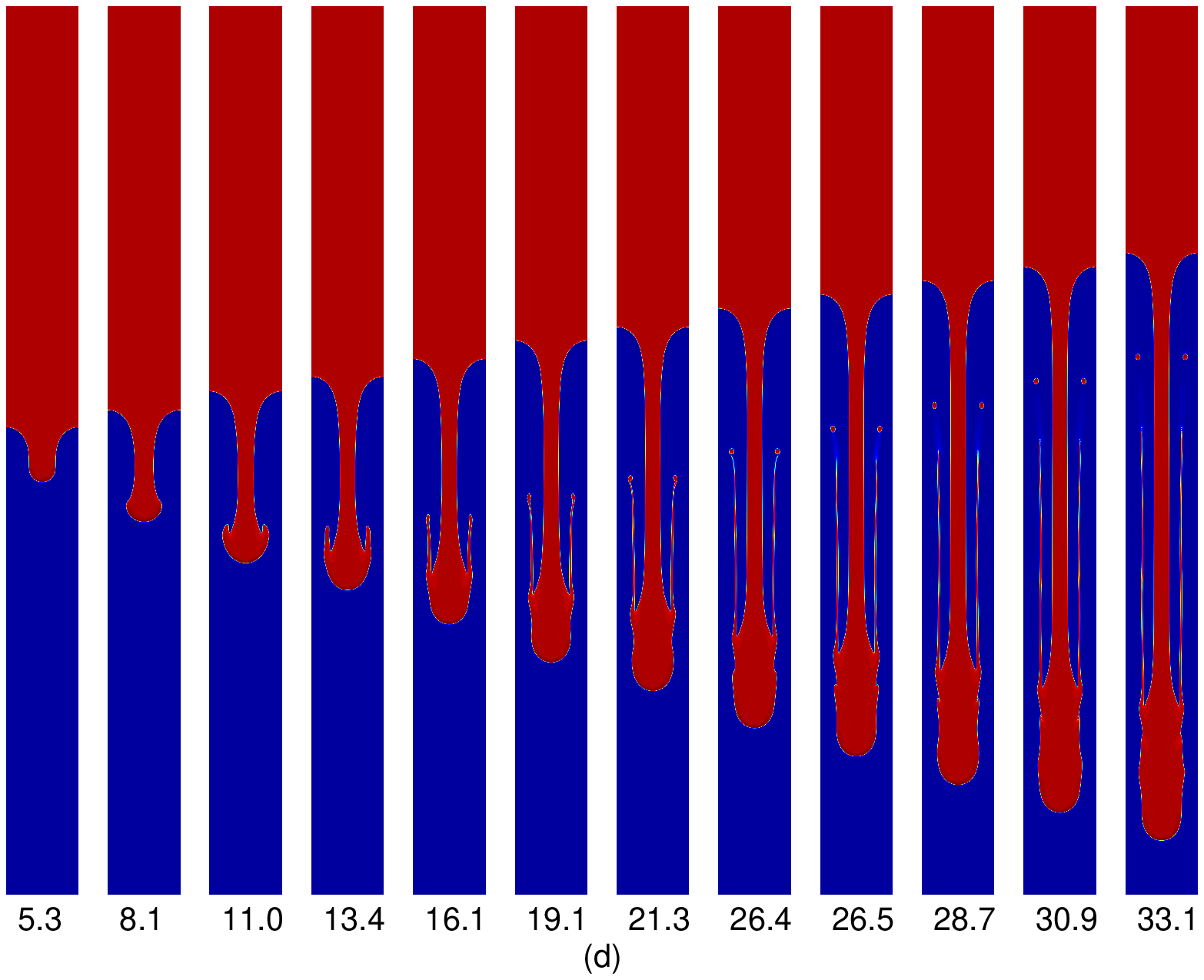} }
\tiny\caption{The time evolutions of the interfacial patterns in the immiscible RTI $(A=0.7)$ with different Reynolds numbers: (a) $Re=10000$, (b) $Re=1000$, (c) $Re=100$, and (d) $Re=10$.}
\end{figure}

The effect of the Reynolds number is first investigated on the late-time dynamics of the single-mode RTI at a moderately high Atwood number of 0.7.
We simulated the evolution of the single-mode instability within a comprehensive range of Reynolds numbers from 1 to 10000 and
the snapshots of interfacial patterns under four typical Reynolds numbers are displayed in Fig. 1.
From this figure, it can be observed that the instability in regardless of various Reynolds numbers exhibits similar interfacial dynamics
at the initial stage: the heavy and light fluids seep into each other forming into the spike and bubble structures. In the following,
the spike and bubble display the distinct behaviours at different Reynolds numbers. For high Reynolds number of 10000, the spike continues to fall
down and the bubble rises up accompanied by the occurrence of the nonlinear effect, which begins to affect the bubble and spike dynamics.
It is seen that the spike rolls up at its front creating a pair of vortices due to the appearance of the nonlinear KH
instability caused by the velocity difference on the bubble-spike interface. Then these two vortices grow in size and
the increasing nonlinear KH instability provides the rolling motion of the interface, resulting in a pair of secondary vortices
at the tails of the roll-ups. As time progresses, the interaction between fluids in the mixing region becomes more and more intensive,
which leads to the roll-ups of the interfaces at the multiple positions and eventually induces the formation of a chaotic interfacial structure
with large topological change. Nonetheless, the close inspection showed that the late-time symmetries of the spike or bubble images at a moderately
high Atwood number can be still preserved familiar from our low-Atwood-number results~\cite{Liang1, Liang2},
which is also qualitatively consistent with the high-resolution direct numerical simulations of miscible single-mode RTI~\cite{Wei,Bian}.
As the Reynolds number is reduced, the roll-up phenomenon appears at a later time and the vortical sizes are accordingly smaller. In general,
the structure of the interface in the evolutional process becomes relatively smooth due to the more stabilized shear layer between bubble and spike. Specially, the heavy fluid at a extremely small Reynolds number of 10 falls down continuously in the manner of the spike almost without the
occurrence of the roll-up or breakup behavior.

\begin{figure}
\subfigure{\includegraphics[width=2.8in,height=2.5in] {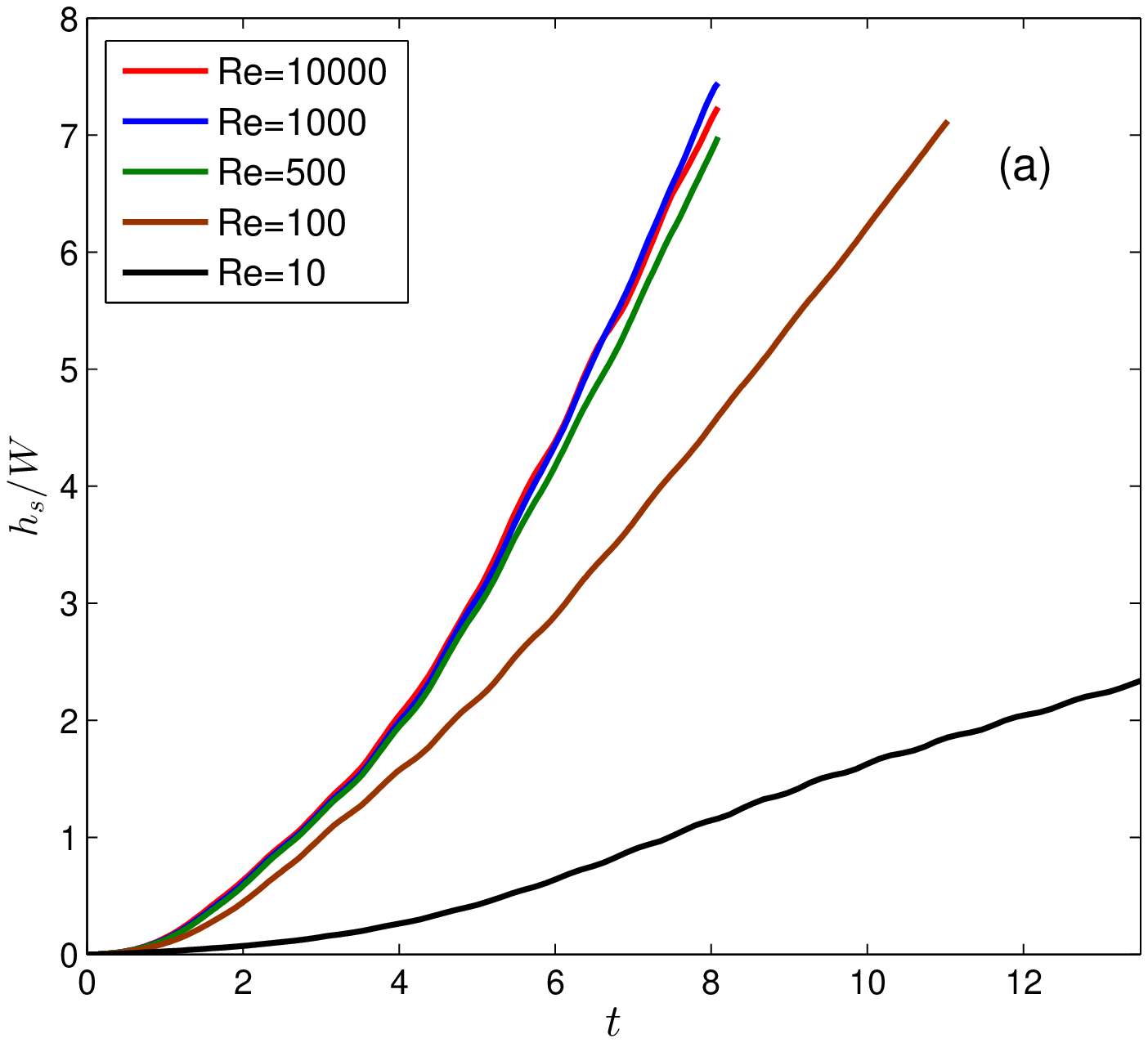} \label{fig:Re-sh}}\hspace{0.5cm}
\subfigure{\includegraphics[width=2.8in,height=2.5in] {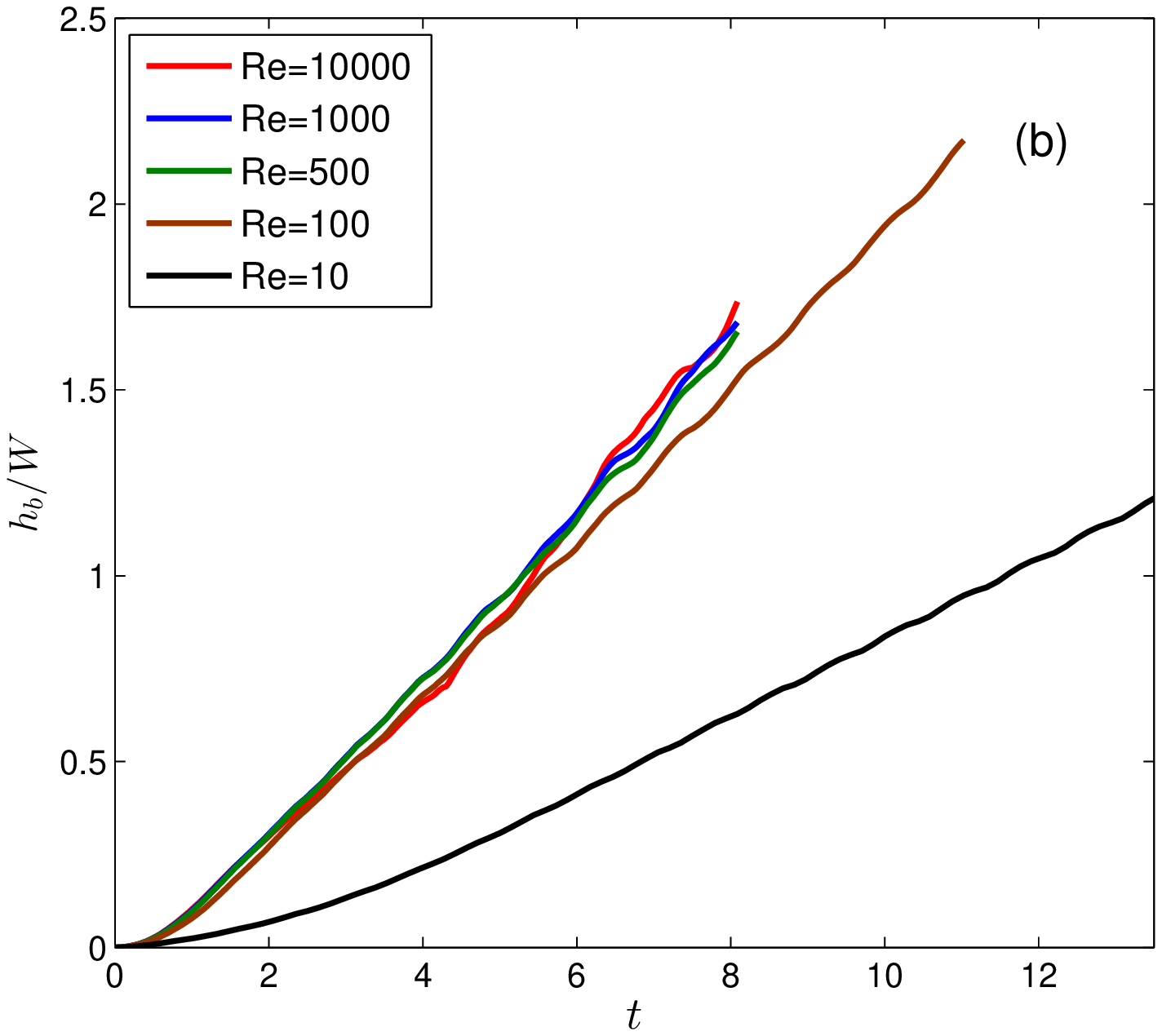} \label{fig:Re-su}}
\tiny\caption{Time variations of (a) normalized spike amplitude and (b) normalized bubble amplitude in the immiscible RTI with various Reynolds numbers.}
\end{figure}

\begin{figure}
\includegraphics[width=2.8in,height=2.5in] {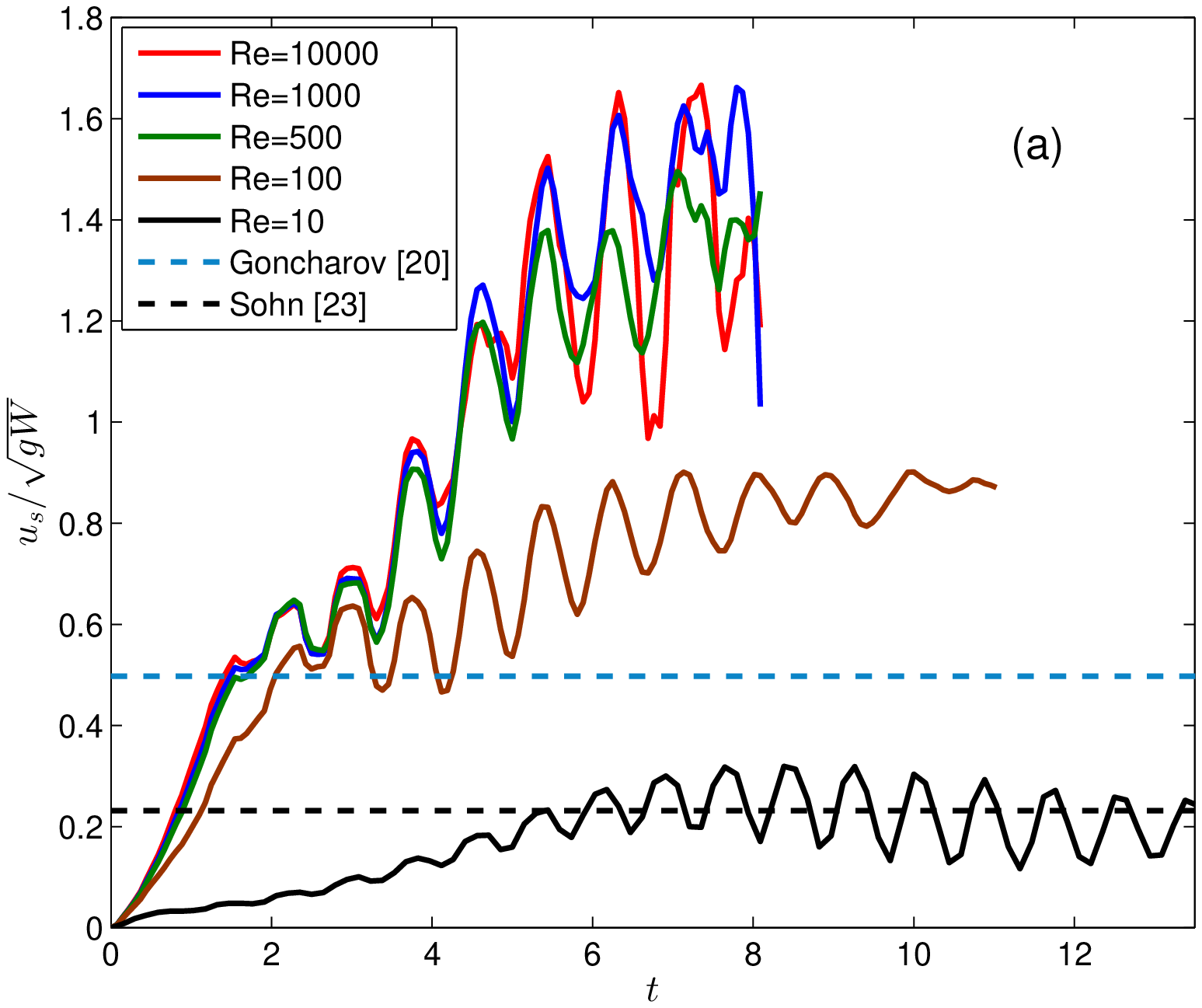}\label{fig:Re-bh}\hspace{0.5cm}
\includegraphics[width=2.8in,height=2.5in] {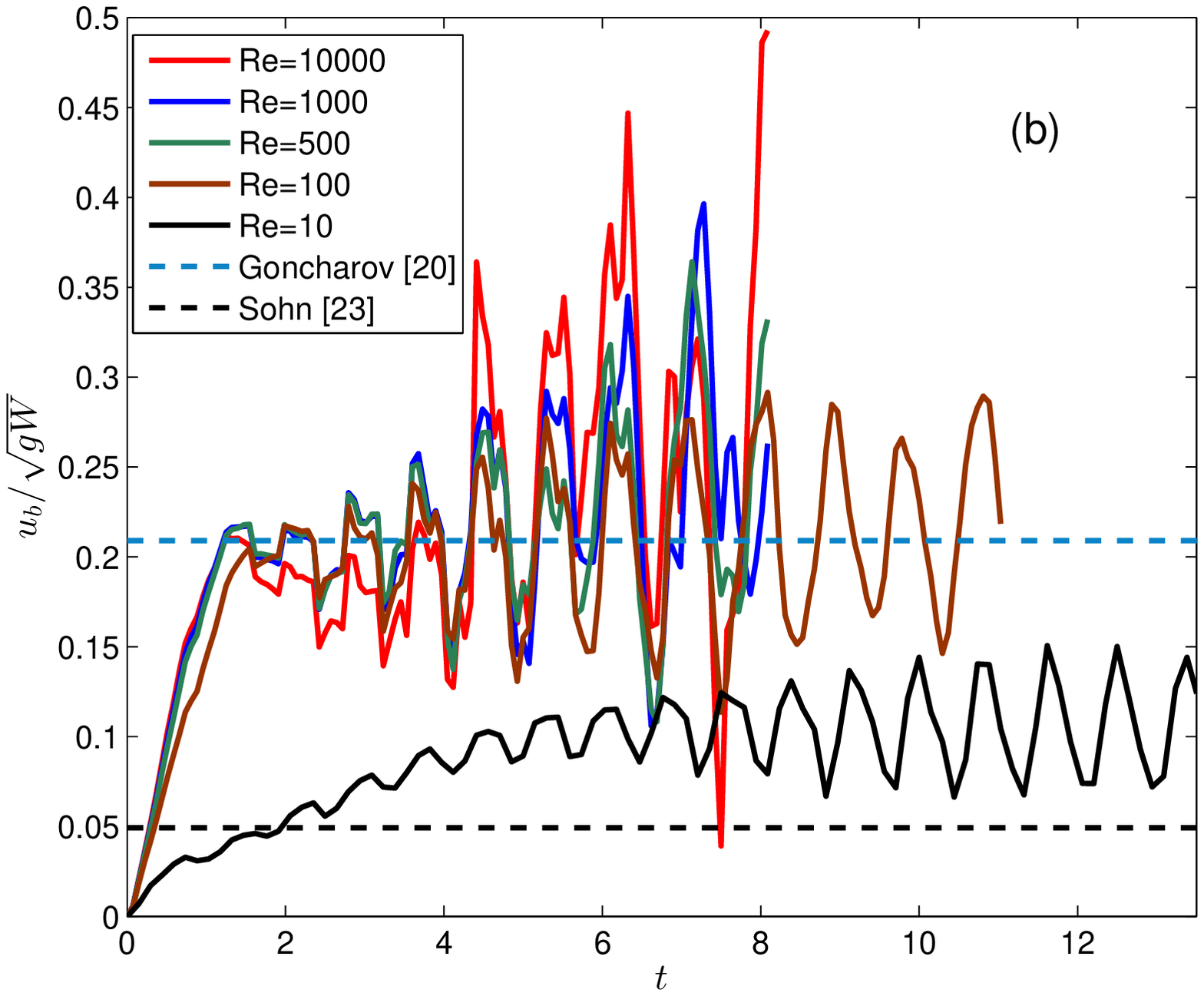}\label{fig:Re-hu}
\tiny\caption{Time variations of (a) normalized spike velocity and (b) normalized bubble velocity in the immiscible RTI with various Reynolds numbers.}
\end{figure}

Further, we quantitatively examined the effect of the Reynolds number on the instability growth in terms of the spike and bubble amplitudes
and velocities, which are often the focuses of the previous studies on the single-mode RTI ~\cite{Waddell, Goncharov, Wilkison, Liang2, Wei}. The spike and bubble amplitudes marked by $h_s$ and $h_b$ are defined as the vertical distances between the spike/bubble fronts and their
corresponding initial positions such that the amplitudes for the spike and bubble at initial time are zero, while their evolutional velocities
denoted by $u_s$ and $u_b$ are calculated from the amplitude profiles by the time derivatives of the amplitudes $h_s$, $h_b$.
Figures 2(a) and 2(b) show the time variations of the spike and bubble normalized amplitudes with a comprehensive range of Reynolds numbers.
It can be found that for all Reynolds numbers, both the spike and bubble amplitudes increase with time and also achieve greater
values at a higher Reynolds number, while the degree of increase is weakened and even the amplitude curves have no obvious differences
when the Reynolds number exceeds a certain value. This indicates that the Reynolds number effect becomes no longer important
when its value is high enough. To further illustrate this statement, we also conduct a theoretical analysis on
the growth of the single-mode instability. The dynamics of the spike front in the incompressible RTI can be described by
a balance per unit mass of the buoyant and dissipation forces~\cite{Abarzhi}
\begin{equation}
\frac{d h_s}{dt}=u_s,~~~~~\frac{d u_s}{dt}=A g+F_d
\end{equation}
where $F_d$ is the dissipation force defined as the rate of momentum loss in the direction of
gravity $F_d=-\varepsilon/\nu$, $\varepsilon$ is the energy dissipation rate. Neglecting the viscous time scale,
the dimensional grounds conducted by Sreenivasan~\cite{Sreenivasan} suggest that the energy dissipation rate can be expressed as
$\varepsilon=C_e\nu^3/W$, where $C_e$ is a positive constant. Substituting the formula into Eq. (25), one can
find that a smaller dissipation force accelerating the spike growth can be achieved in theory for a larger Reynolds number.
Furthermore, when the Reynolds number is sufficiently large, the dissipation force becomes extremely small, in this case the
gravitational force dominates over the dissipation force such that the change in fluid viscosity has no significant influence
on the spike development. This theoretical analysis can be well confirmed in our numerical experiments,
where the spike and bubble amplitudes indeed increase with the Reynolds number at first and then almost are independence
of the sufficiently high Reynolds number. In addition, the comparison between the spike and bubble amplitude curves implies
that the spike front grows more quickly than that of the bubble. The time evolutions of the spike and bubble velocities with
these typical Reynolds numbers are also measured in Figs. 3(a) and 3(b). From these figures, we identified the distinct growth
stages as describing the evolution of single-mode instability with a moderately large Atwood number at different Reynolds numbers.
For high or medium Reynolds numbers, the development of the single-mode RTI experiences four distinguishing stages named as the linear growth,
saturated velocity growth, reacceleration, and chaotic growth stages. At the linear stage, the perturbation grows with an
exponential function~\cite{Waddell} followed by the saturated velocity growth stage, where the bubble and spike evolve with constant velocities,
as can be clearly seen in Figs. 3(a) and 3(b). For comparisons, the analytical potential flow model by Goncharov~\cite{Goncharov}
for predicting the spike and bubble saturated velocities given in Eq. (1) is also presented. It can be found that the present numerical
results quantitatively agree well with the corresponding analytical solutions. Besides, the duration of the saturated velocity growth
stage for the spike in the relatively high-Atwood-number RTI is shown to be transient compared with the case of the low-Atwood-number~\cite{Liang1, Wei}, which was also observed in the numerical simulation of single-mode RTI with a moderate Atwood number~\cite{Goncharov}. As for the bubble, its lasting time of this stage is found to be longer than that of the spike. After the saturated velocity stage, the increasing nonlinear strength of vortices in the proximities of the spike and bubble tips accelerates the growths of the spike and bubble, making their evolutional velocities exceed the values of the potential flow theory~\cite{Goncharov}, which indicates that the instability has entered into the reacceleration stage. The reacceleration stage cannot last indefinitely. From Figs. 3(a) and 3(b), it can be shown that the spike and bubble velocities at the late time become unstable and are accelerated and decelerated repeatedly, suggesting the evolution of the stability has been transformed into the chaotic development stage. The chaotic stage was primarily found by Ramaprabhu {\it{et al.}}~\cite{Ramaprabhu1} using the integrated large eddy simulation, while different from the present observation, they reported that the spike and bubble velocities at the chaotic stage showed an decrease with time probably owning to the low resolution of the adopted computational grid. Actually, the fluctuation behaviors for the spike and bubble velocities were also observed in the high-resolution direct numerical simulations~\cite{Wei,Hu1, Bian} and our mesoscopic lattice Boltzmann simulations~\cite{Liang1, Liang2} for low Atwood number fluid system. As the Reynolds number is gradually reduced, we can observe from Figs. 3(a) and 3(b) that some later stages such as the reacceleration and chaotic development stages cannot be reached successively. For instance, the bubble and spike evolutional velocities at a low Reynolds number of 10 are observed to have initially slight fluctuations at the late time and eventually approach the quasisteady values without the appearances of the reacceleration and chaotic stages. In addition, it is noted that the spike and bubble quasisteady velocites are smaller than the solutions of the potential flow theory owing to the neglected viscosity effect in his analysis~\cite{Goncharov}. Sohn~\cite{Sohn} realized this problem and modified the asymptotic bubble velocity including the viscosity and surface tension force effects as
\begin{equation}
u_b=\sqrt{\frac{2Ag}{3(1+A)k}-\frac{k\sigma}{9\rho_h}+\frac{4}{9}k^2\nu^2}-\frac{2}{3}k\nu,
\end{equation}
and the corresponding spike asymptotic velocity inspired by the work of Goncharov~\cite{Goncharov} can be written as
\begin{equation}
u_s=\sqrt{\frac{2Ag}{3(1-A)k}-\frac{k\sigma}{9\rho_h}+\frac{4}{9}k^2\nu^2}-\frac{2}{3}k\nu.
\end{equation}
We also plotted in Figs. 3(a) and 3(b) the above theoretical solutions of the spike and bubble asymptotic velocities at
a low Reynolds number of 10. It can be found that the numerical prediction of the spike velocity at the saturated velocity stage
agrees well with the theoretical solution, while the numerical result for the bubble saturated velocity is higher than
the corresponding theoretical solution. This discrepancy may be caused by the effect of some inapparent vortices~\cite{Betti} and needs
to be deeply investigated in future research.

\begin{figure}
\includegraphics[width=2.8in,height=2.5in] {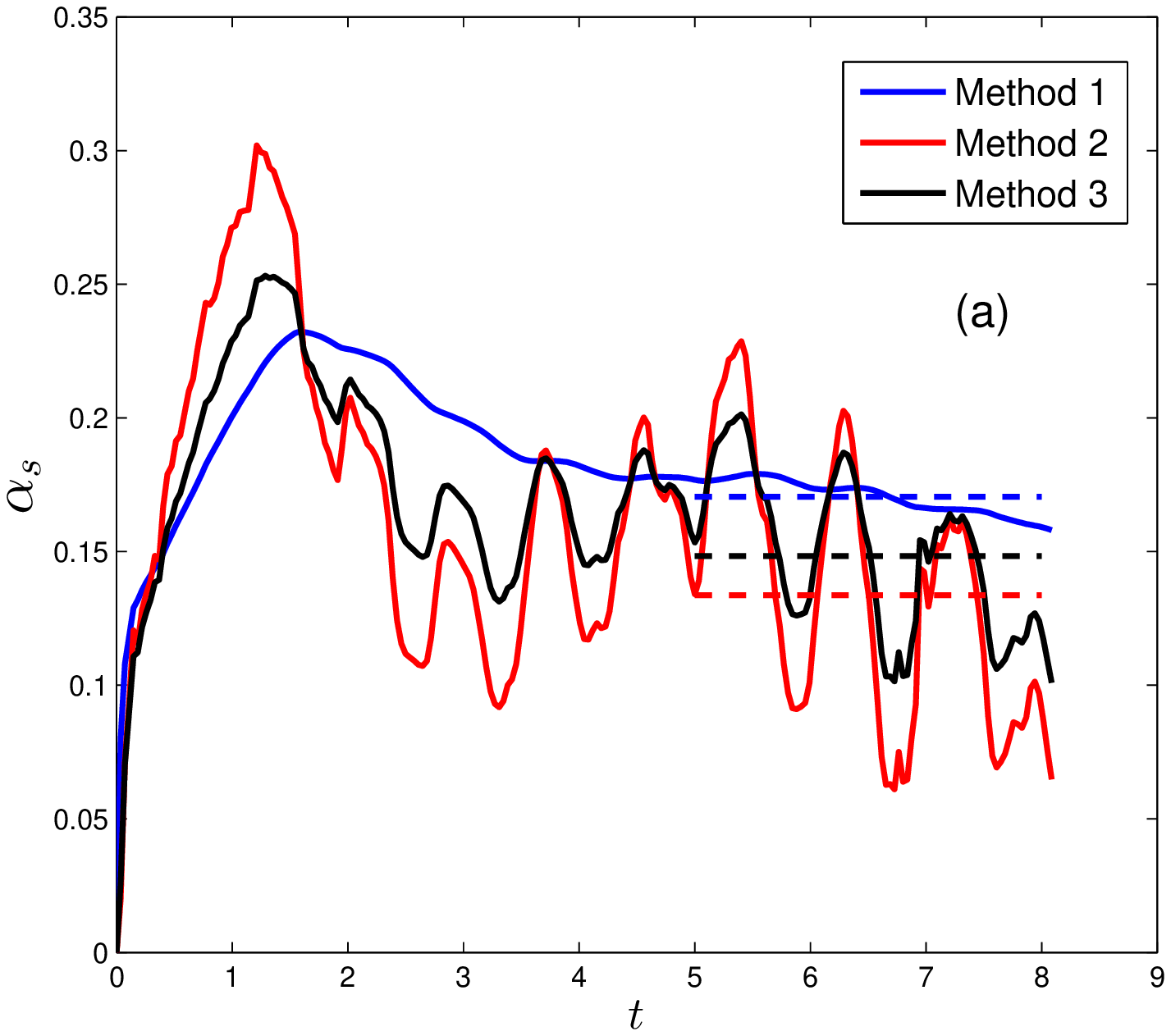}\hspace{0.5cm}
\includegraphics[width=2.8in,height=2.5in] {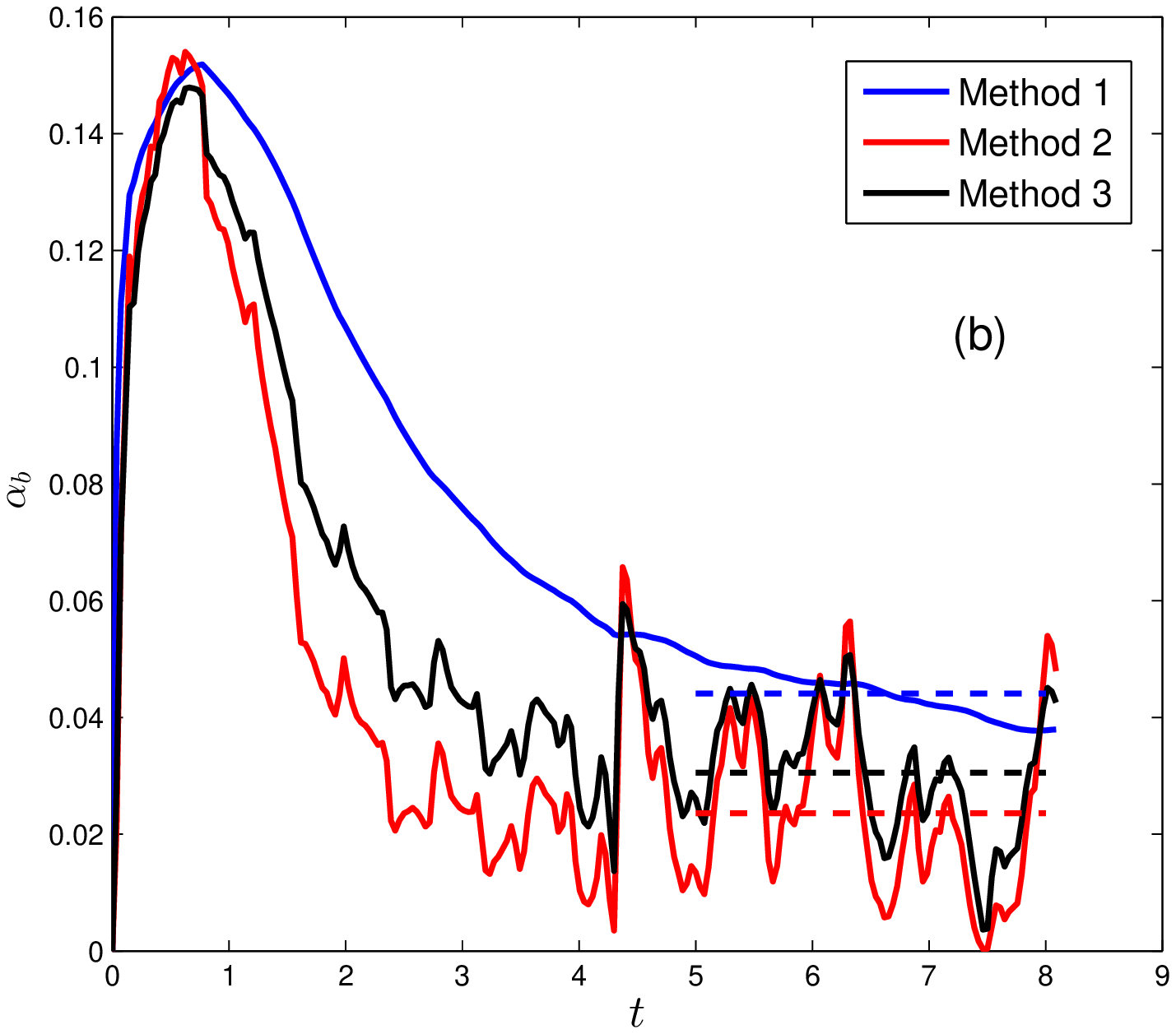}
\tiny\caption{Time variations of (a) spike and (b) bubble normalized accelerations computed by Methods 1, 2 and 3. The dotted lines represent the averages of the spike and bubble accelerations at the later period.}
\end{figure}	
	
\begin{figure}
\subfigure{\includegraphics[width=2.8in,height=2.5in] {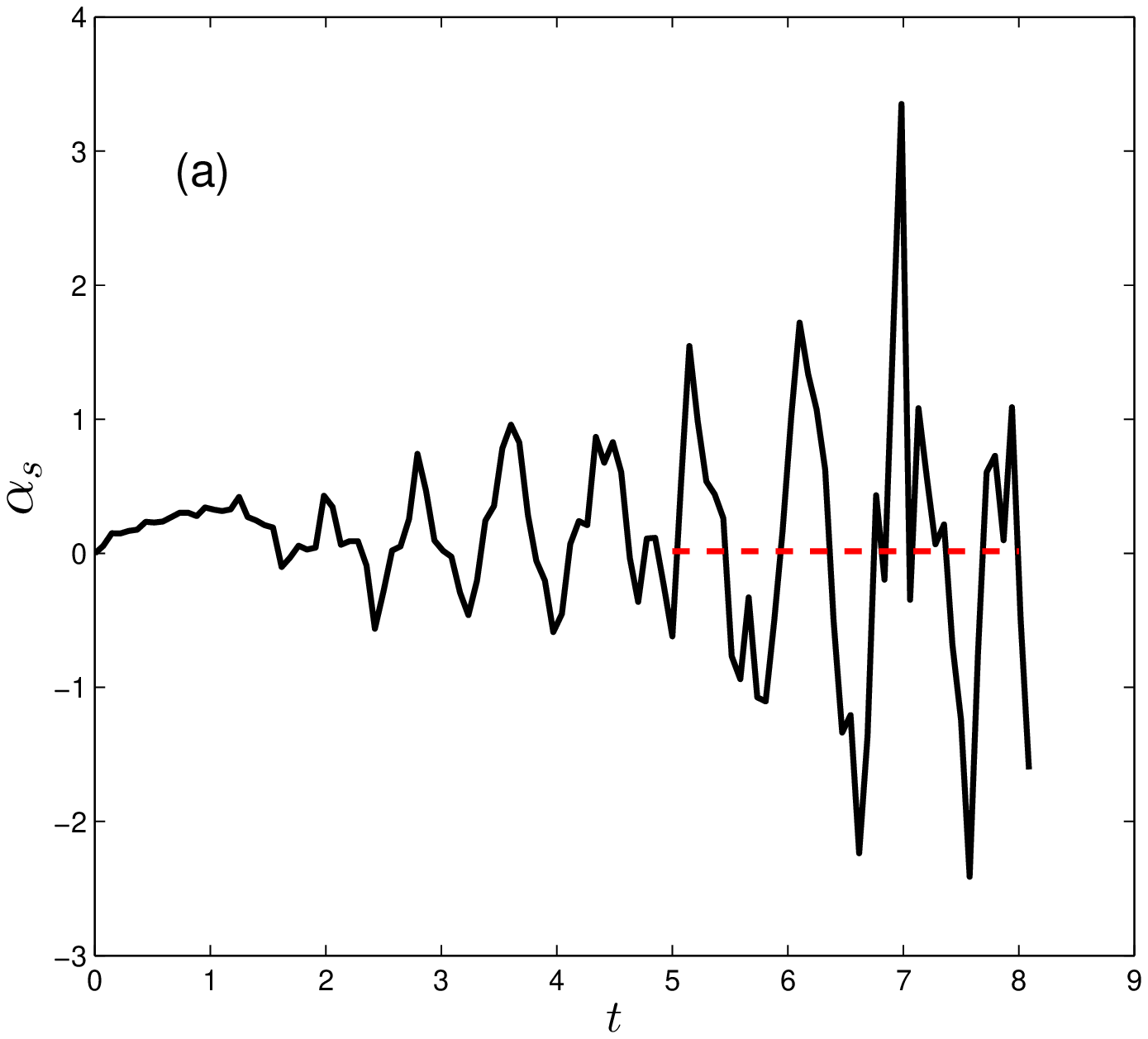}}\hspace{0.5cm}
\subfigure{\includegraphics[width=2.8in,height=2.5in] {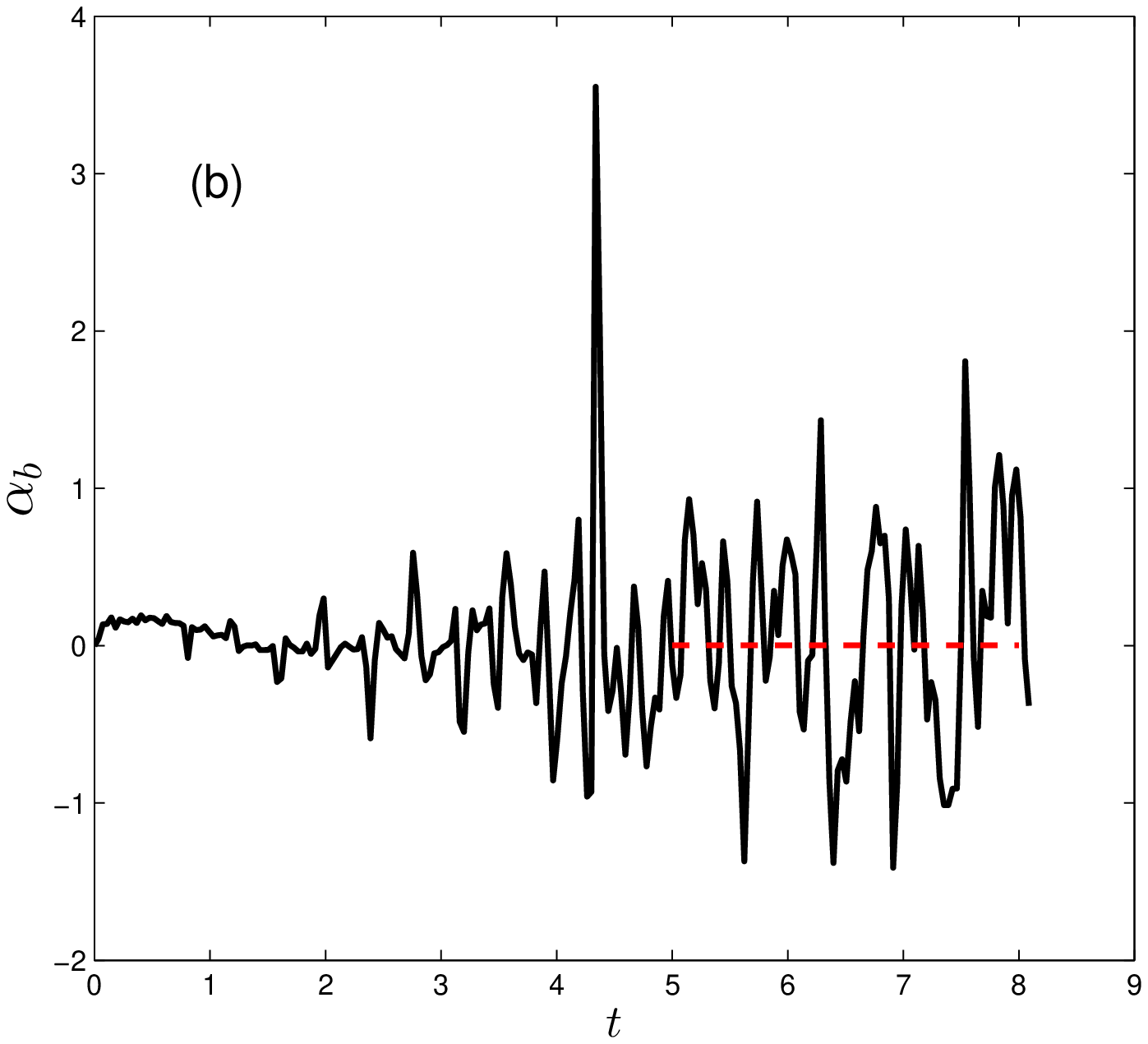}}\hfill
\subfigure{\includegraphics[width=2.8in,height=2.5in] {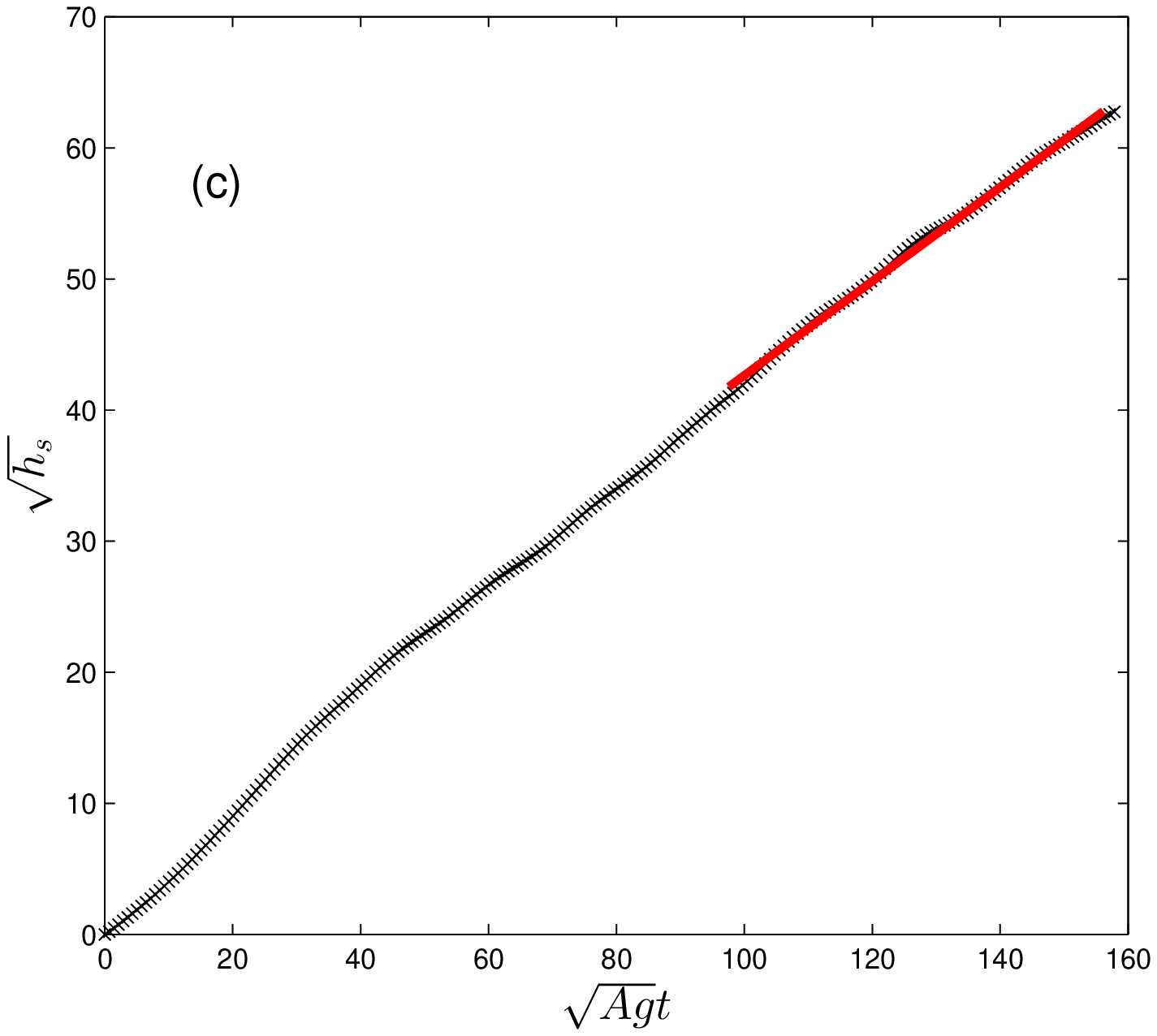}}\hspace{0.5cm}
\subfigure{\includegraphics[width=2.8in,height=2.5in] {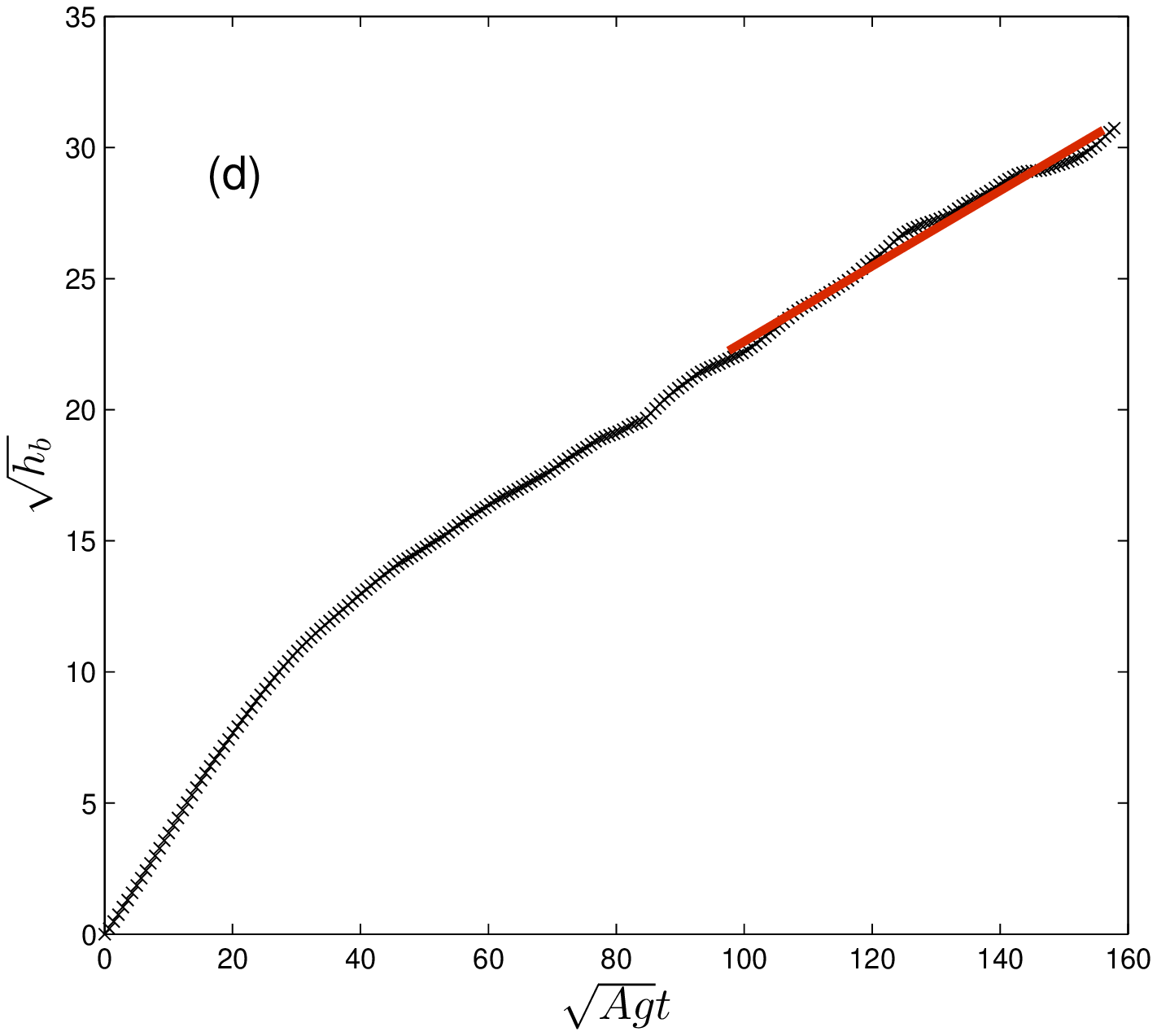}}
\tiny\caption{Time variations of (a) spike and (b) bubble normalized accelerations computed by Method 4 and the dotted lines are the averaging values at the later stage. The dependence of (c) $\sqrt{h_{s}}$ and (d) $\sqrt{h_{b}}$ on the variable $\sqrt{Ag}t$ by Method 5 and the solid lines represent their late-time linear fitting curves.}
\end{figure}
	
Based on the latest knowledge~\cite{Wei, Liang2, Bian}, it has been recognized that the late-time growth of the single-mode RTI with high
Reynolds numbers exhibits a mean quadratic law in the terms of
\begin{equation}
h_{s,b}=\alpha_{s, b }g A t^2,
\end{equation}
where $h_{s, b }$ with the subscripts $s$ and $b$ represent the spike and bubble amplitudes, and $\alpha_{s, b }$ are the corresponding spike and
bubble growth rates. To determine the late-time growth nature of the single-mode RTI with a moderately high Atwood number, we also measured
the spike and bubble growth rates at a high Reynolds number of 10000. To data, there have been several statistical approaches to compute
the growth rate coefficients of the multi-mode phenomenon, which will be adopted here to derive the spike and bubble late-time growth
rates of single-mode instability and the comprehensive comparisons of these statistical methods are also conducted. Based on their basic
definitions given in Eq. (28), the coefficients reflecting the spike and bubble growth rates can be calculated via the following relation~\cite{Cabot}
\begin{equation}
{\alpha_{s,b}}={\frac{h_{s,b}}{Agt^2}}.
\end{equation}
Ristorcelli {\it{et al.}}~\cite{Ristorcelli} and Cook {\it{ et al.}}~\cite{Cook} using completely different approaches,
deduced the same formula for computing the spike and bubble growth rates, which can be expressed as
\begin{equation}
\alpha_{s,b}={\frac{{{\dot{h}}_{s,b}}^2}{4Agh}},
\end{equation}
where ${\dot{h}}_{s,b}$ with the subscripts $s$ and $b$ respectively denote the time derivatives of the spike and bubble amplitudes.
To achieve a better convergence, Clark~\cite{Clark} also proposed another measurement method to obtain
the spike and bubble growth rates,
\begin{equation}
{\alpha_{s,b}}={\frac{\partial{h_{s,b}}}{\partial{Z}}}={\frac{\partial{h_{s,b}}}{\partial{z}}}{\frac{\partial{z}}{\partial{Z}}}=
{\frac{1}{2A} \frac{\partial{h_{s,b}}}{\partial{z}}},
\end{equation}
where $Z=2Az$, $z={gt^2/2}$, $t$ is the time that the acceleration is applied. In addition, Wei {\it{et al.}}~\cite{Wei} presented a
direct statistical manner to determine the spike and bubble growth rates,
\begin{equation}
\alpha_{s,b}={\frac{{{\ddot{h}}_{s,b}}}{2Ag}},
\end{equation}
where ${{\ddot{h}}_{s,b}}$ are the second derivatives of the spike and bubble amplitudes with respect to the evolutional time.
Recently, Olson and Jacobs~\cite{Olson} took the square root of Eq. (28) yielding a linear function in time,
\begin{equation}
h^{1/2}_{s,b}={({\alpha_{s,b}}Ag)^{1/2}}t,
\end{equation}
thus the coefficients $\alpha_{s,b}$ can be easily extracted by plotting the linear fitting curves of $h^{1/2}_{s,b}$
versus $(Ag)^{1/2}t$ at the late time. Most of these statistical methods are proposed in order to calculate the growth coefficients of
the multi-mode instability with an exception~\cite{Wei}. Although they seem to be significantly different in form, they are mathematically
equivalent but could produce distinct behaviours of growth rate curves in practical applications~\cite{Cabot, Akula}. For the convenience of the discussion, the aforementioned five statistical methods given in Eq. (29) to (33) are orderly labeled as Method 1 to 5. Figure 4 depicts the time evolutions of the spike and bubble accelerations computed directly from Methods 1, 2 and 3. It can be seen that the spike and bubble normalized
accelerations by Method 1 approximate to nearly constant values with slight reductions by the end of the simulation, while the
predictions of the spike and bubble accelerations by Methods 2 and 3 have some fluctuations at the late-time stage. We conducted the
late-time averages of these acceleration curves to obtain the spike and bubble growth rates, and the measured results are summarized in Table 1.
It is shown that the spike growth rates $\alpha_s$ by Methods 1, 2 and 3 are 0.1705, 0.1337 and 0.1483 respectively, while the corresponding
bubble growth rates $\alpha_b$ are 0.0441, 0.0236 and 0.0305. Further, we also presented the snapshots of the spike and bubble standardized
accelerations obtained by Method 4 in Figs. 5(a) and 5(b), and find that the growth rates fluctuate around the mean values of 0.0157 and 0.0017,
respectively. In addition, the relations between $\sqrt{h_{s,b}}$ and $\sqrt{Ag}t$ are plotted in Figs. 5(c) and 5(d), where the linear
fitting curves at the late time are also presented, indicating the measured spike and bubble growth rates $\alpha_s$ and $\alpha_b$ as
0.1289 and 0.0207, respectively. Though the comparisons of these predicted growth rates in Table 1, it can be found that the result
of Method 4 is far less than those of other four statistical methods, which thus is not a good choice in the measurements of the late-time
growth rates. Meanwhile, the largest values for the growth rates achieved by Method 1 would also be not applied, and the prediction of the
bubble growth rate by Method 3 is relatively high. In general, the results of Methods 2 and 5 are close to each other and will be recommended in
the following calculations of the spike and bubble growth rates with various Atwood numbers. As we noted, it has been reported that
Methods 2 and 5 also produced the comparative results in predicting the growth rates of multi-mode RTI and would
be the preferred choices~\cite{Akula}.

\begin{table*}
\caption{\label{table1}The measured spike and bubble growth rates $\alpha_s$ and $\alpha_b$ using five statistical methods}
\begin{tabular}{ccc}
\hline
Statistical method ~~~~~~& $\alpha_s$ ~~~~~~& $\alpha_b$ \\
 \hline
Method 1  & 0.1705 ~~~& 0.0441~~~  \\
Method 2  & 0.1337 ~~~& 0.0236~~~  \\
Method 3  & 0.1483 ~~~& 0.0305~~~  \\
Method 4  & 0.0157 ~~~& 0.0017~~~  \\
Method 5  & 0.1289 ~~~& 0.0207~~~  \\
\hline
\end{tabular}
\end{table*}

\subsection{Effect of the Atwood number}

\begin{figure}
\subfigure{\includegraphics[width=2.9in,height=2.8in] {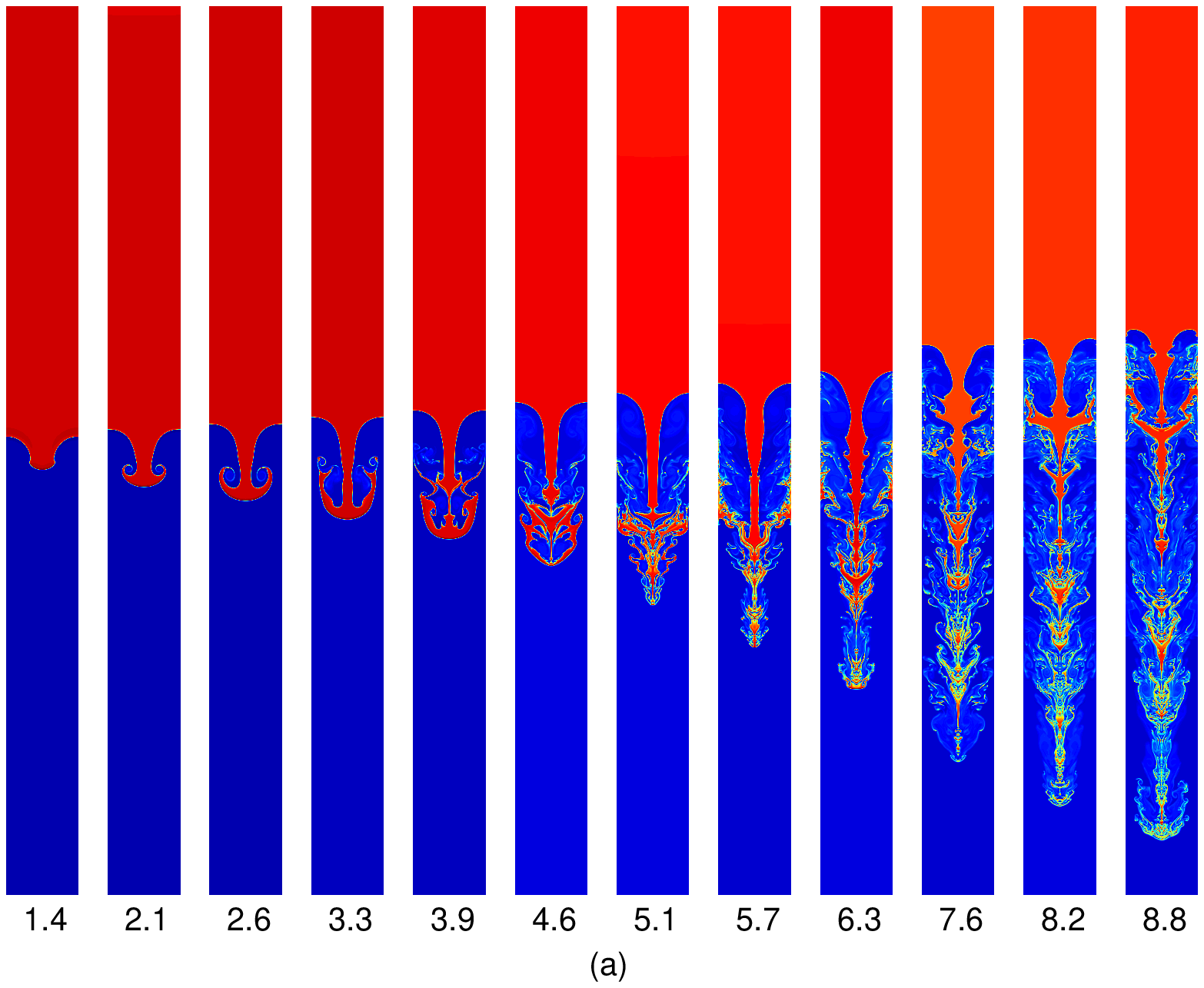} }\hspace{0.7cm}
\subfigure{\includegraphics[width=2.9in,height=2.8in] {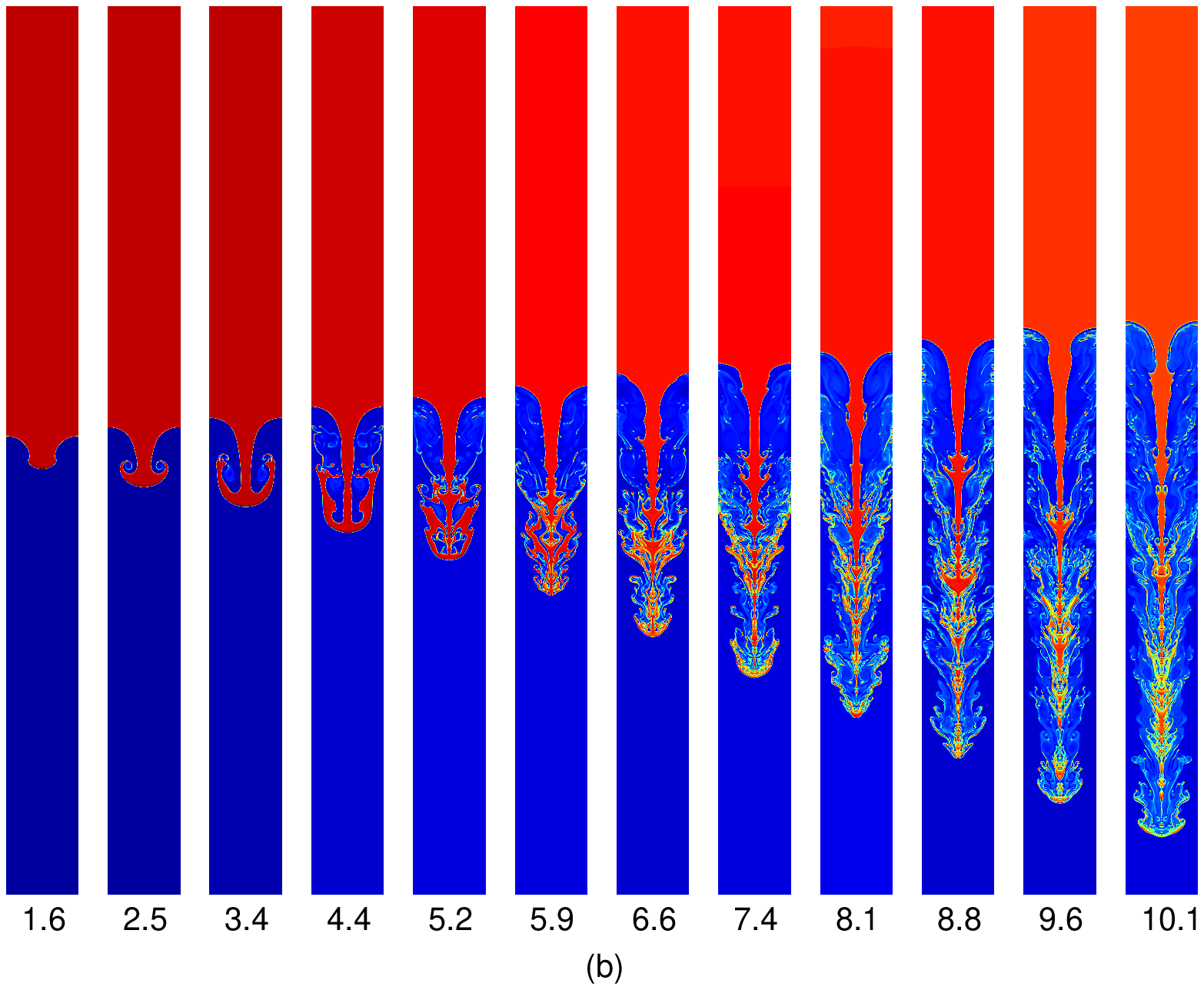}}\vskip 16pt
\subfigure{\includegraphics[width=2.9in,height=2.8in] {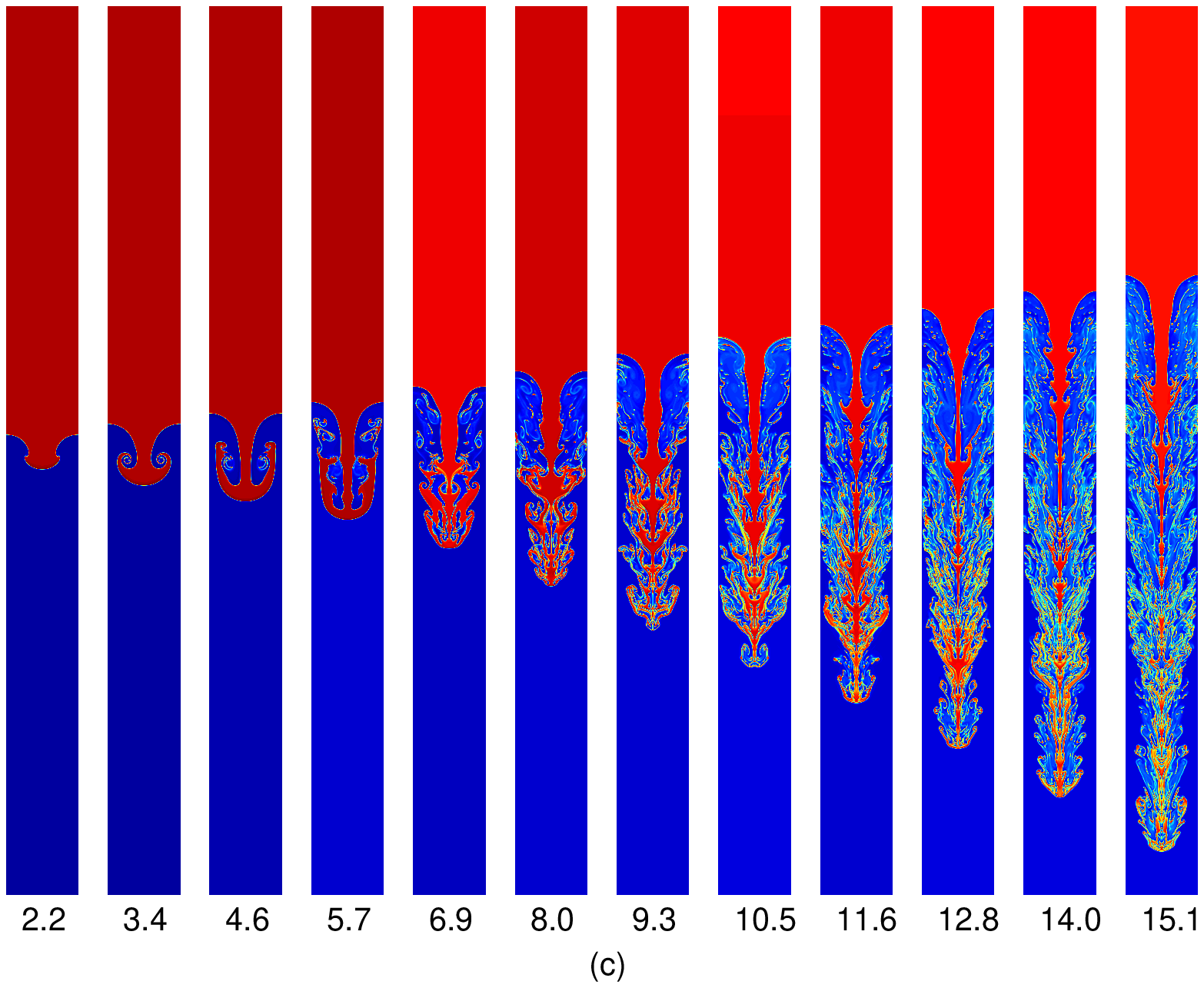} }\hspace{0.7cm}
\subfigure{\includegraphics[width=2.9in,height=2.8in] {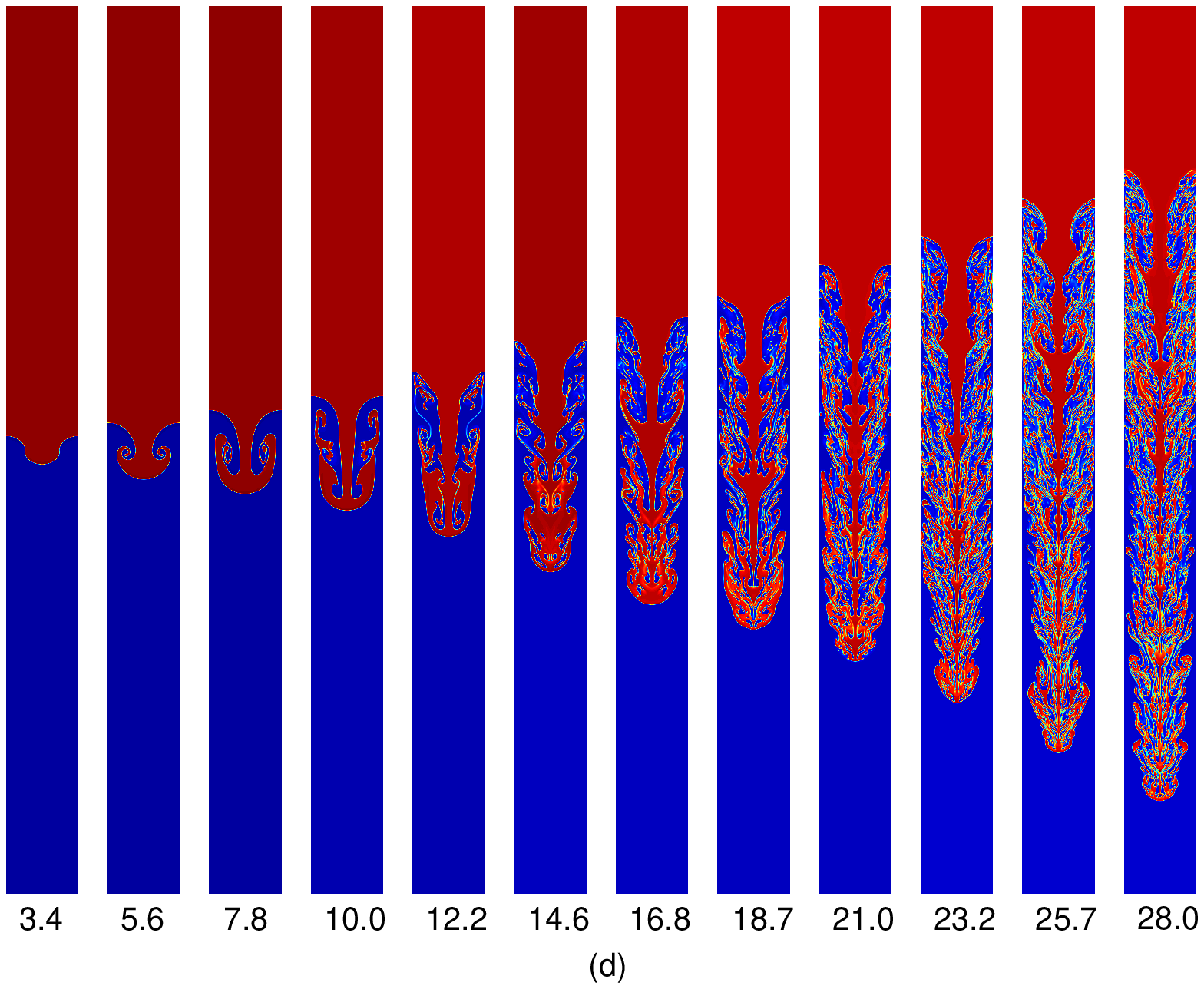} }
\tiny\caption{Time evolutions of the interfacial patterns in the immiscible RTI $(Re=10000)$ with different Atwood numbers: (a) $A=0.6$, (b) $A=0.5$, (c) $A=0.3$, and (d) $A=0.1$.}
\end{figure}

\begin{figure}
\includegraphics[width=2.8in,height=2.5in] {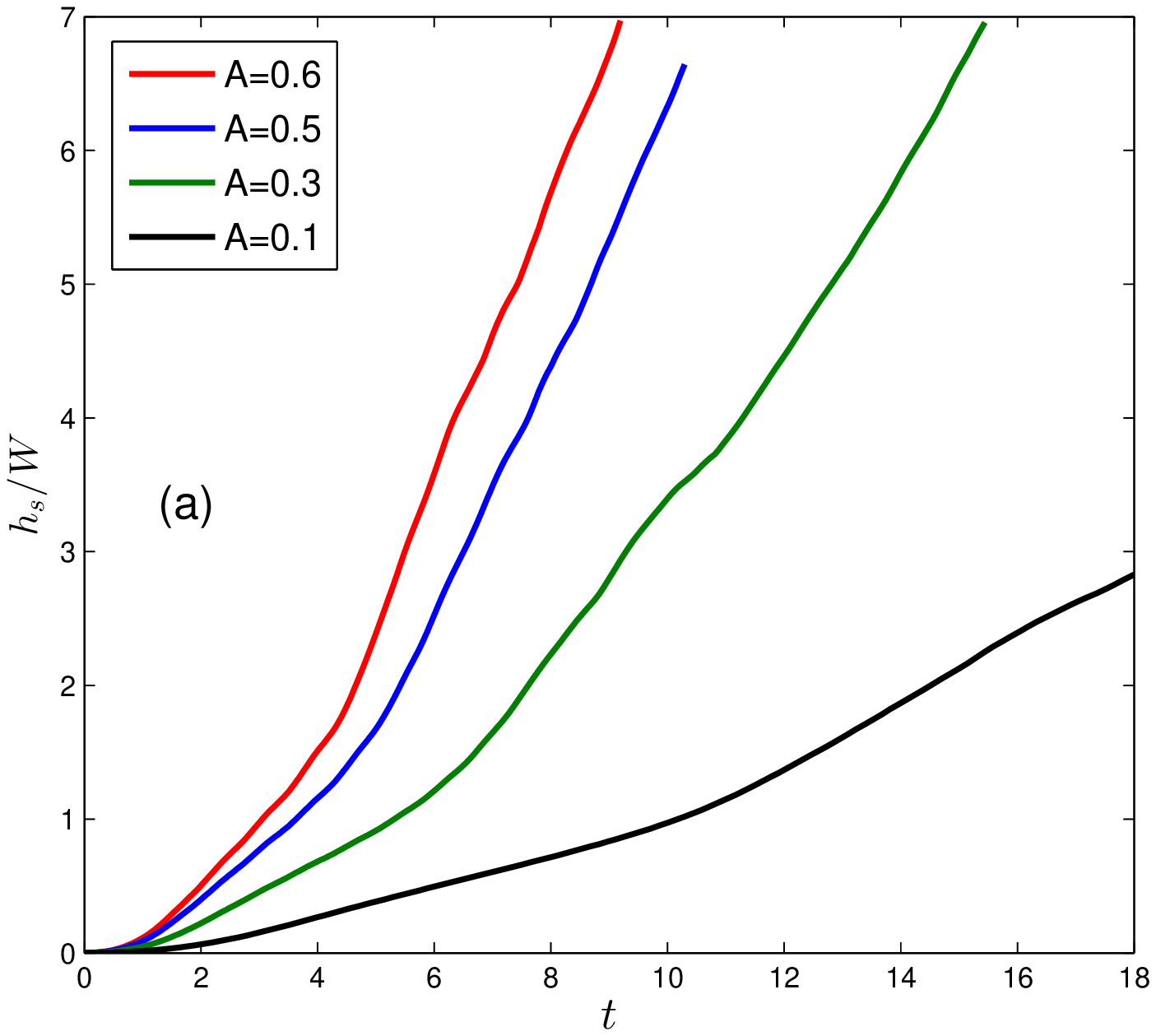}\hspace{0.5cm}
\includegraphics[width=2.8in,height=2.5in] {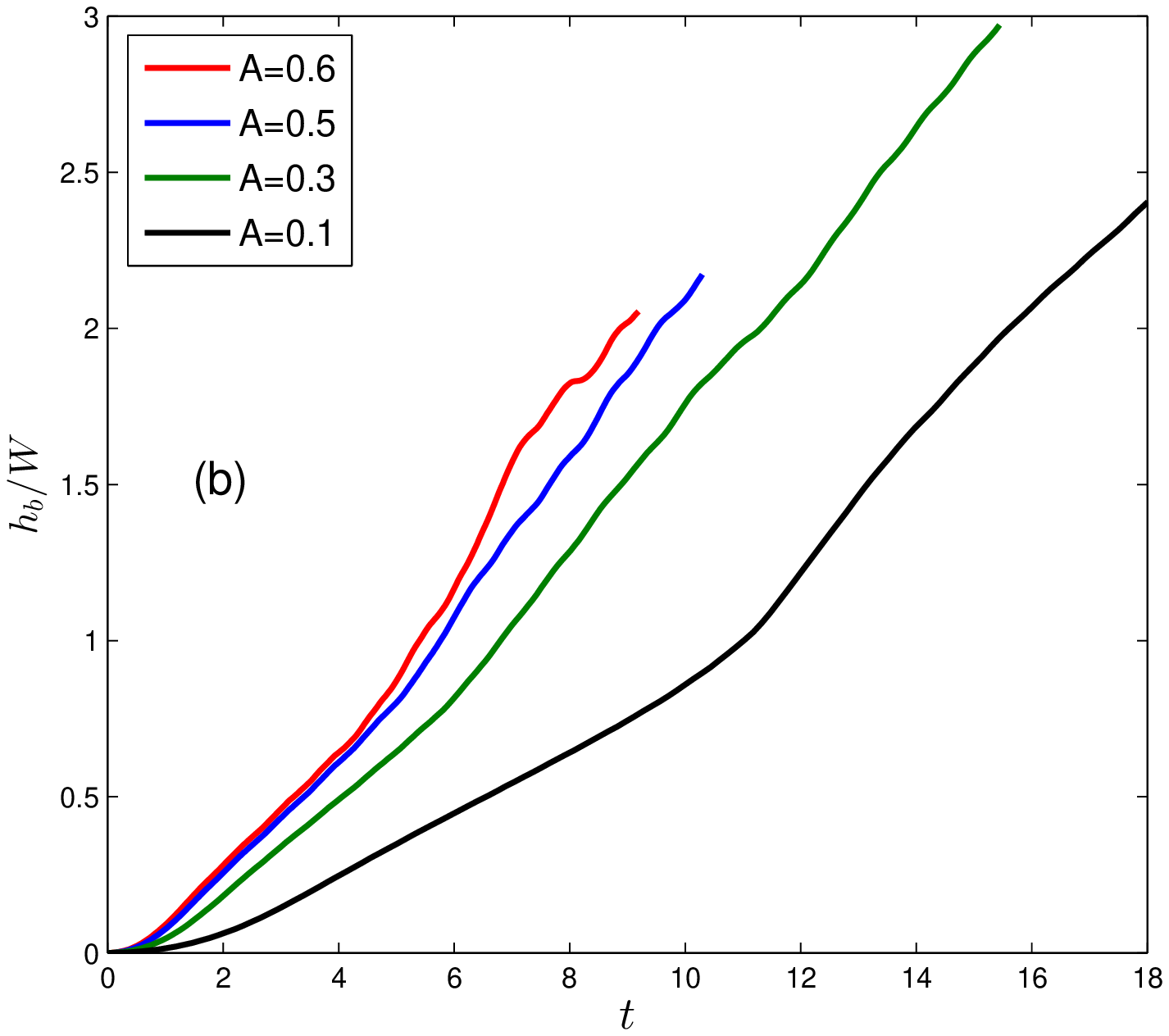}
\tiny\caption{Time variations of (a) normalized spike amplitude and (b) normalized bubble amplitude in the immiscible RTI with various Atwood numbers.}
\end{figure}

In this subsection, we will conduct a late-time description of single-mode RTI with an extensive range of Atwood numbers under the circumstance of a sufficiently high Reynolds number of 10000. Figure 6 depicts the time evolutions of the density visualizations at four different Atwood numbers. It can be found that for all Atwood numbers, a heavy fluid and a light one penetrate into each other at the early stage and form into
the spike and bubble, respectively. The spike then rolls up along its flank and a pair of vortices are visible in the system. The size of the
vortex is shown to decrease with the Atwood number. The portion of the roll-ups further develops and becomes much elongated. Later, a pair of secondary vortices can be generated at the tail of the roll-ups. As time goes on, the interactions between fluids are gradually strengthened owning to the increasing KH instability, which provides the rolling motion of the interfaces at multiple positions and a complex large-scale interfacial structure can be formed. Finally, the large-scale structure continues to increase in size until the mixing layer becomes fully turbulent. Meanwhile, the interfaces in the mixing layer undergo dramatic deformations and some of them even have a chaotic breakup, inducing the formation
of numerous dissociative droplets in the system. The glance of Fig. 6 showed that for all Atwood numbers, the spike and bubble patterns in the whole evolutional process can roughly preserve the symmetric properties with respect to the middle vertical line. However, it is also observed that the spike falls down more quickly and the light fluid occupies a less space of the mixing region as the Atwood number increases. The asymmetric development between the spike and bubble becomes more significant at a higher Atwood number. For example, the spike amplitude $h_s$ is about 3.0 times larger than the bubble amplitude $h_b$ for $A=0.6$ at $t\approx3.0$, while the relation $h_s\approx1.2h_b$ is achieved for $A=0.1$ when the spike reaches the same position. This can be clearly seen in Figs. 7(a) and 7(b), where the time variations of the spike and bubble amplitudes with different Atwood numbers are presented. We can further observe from these figures that the amplitude of the spike increases with time but achieves a greater value at a higher Atwood number, indicating in this case the growth of the spike is accelerated by enlarging the Atwood number and that was also reported in other numerical method simulations of single-mode RTI~\cite{Ramaprabhu1, Bian, Luo}. Besides, the bubble amplitude is found to follow the similar rule that increases with the Atwood number, while the increase range is reduced as the Atwood number is large enough. We would like to point out that the present result on the bubble growth seems to be contradictory with the previous report on the single-mode RTI~\cite{Ramaprabhu1, Bian, Luo}, where the bubble amplitude under the incompressible condition shows a trend of reduction with the Atwood number. This inconsistency can be attributed to the adopted different definitions on the characteristic time. We define the characteristic time in this paper no longer containing the Atwood number, while other articles generally adopt the characteristic time relating to the Atwood number as $\sqrt{\frac{(1+A)W}{Ag}}$~\cite{Ramaprabhu1, Bian, Luo}, leading to a larger value at a lower Atwood number such that it needs a longer iteration step for the flow to reach the same dimensionless time and thus achieves a larger bubble amplitude.
	
\begin{figure}
\includegraphics[width=2.8in,height=2.5in] {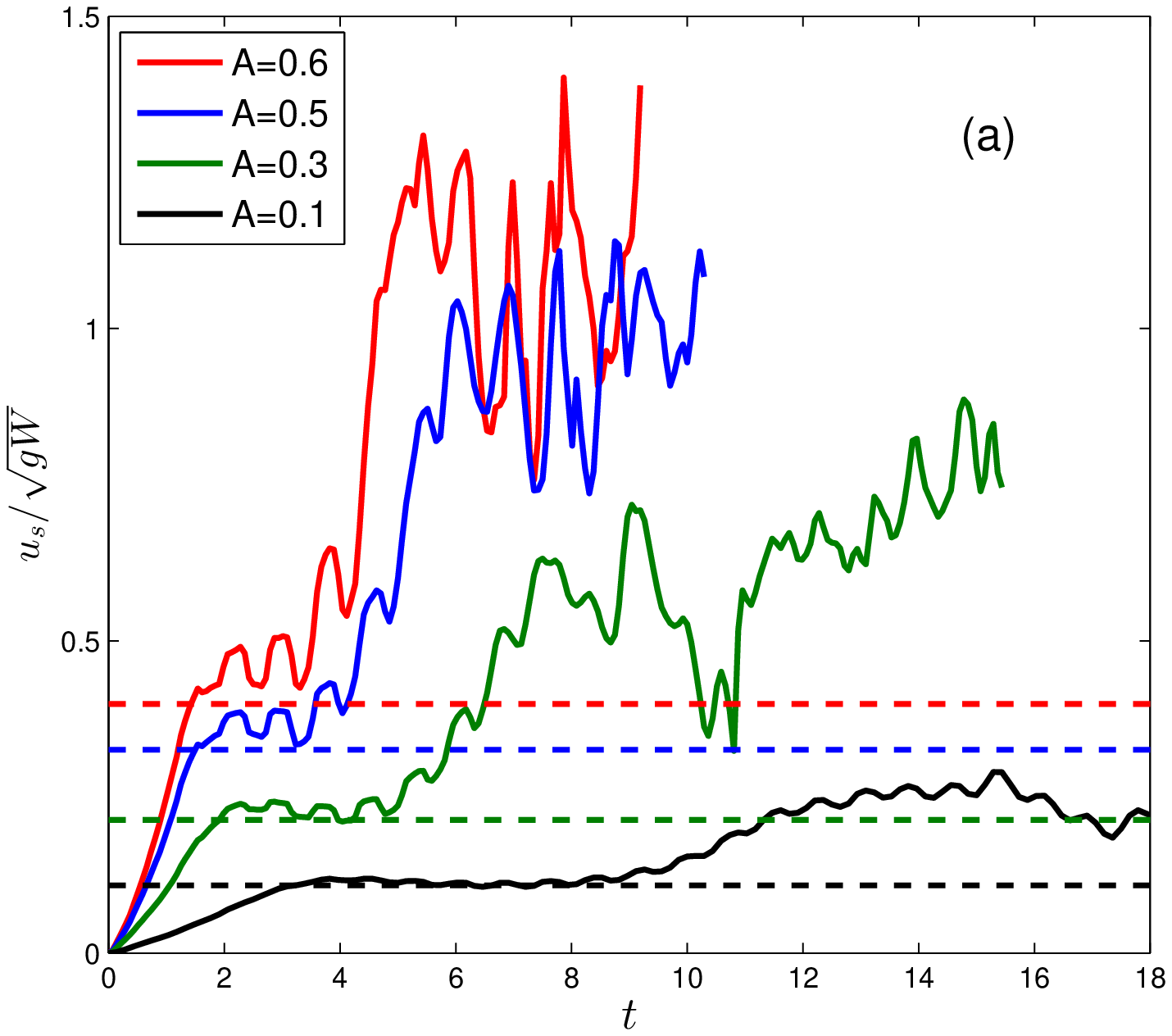}\hspace{0.5cm}
\includegraphics[width=2.8in,height=2.5in] {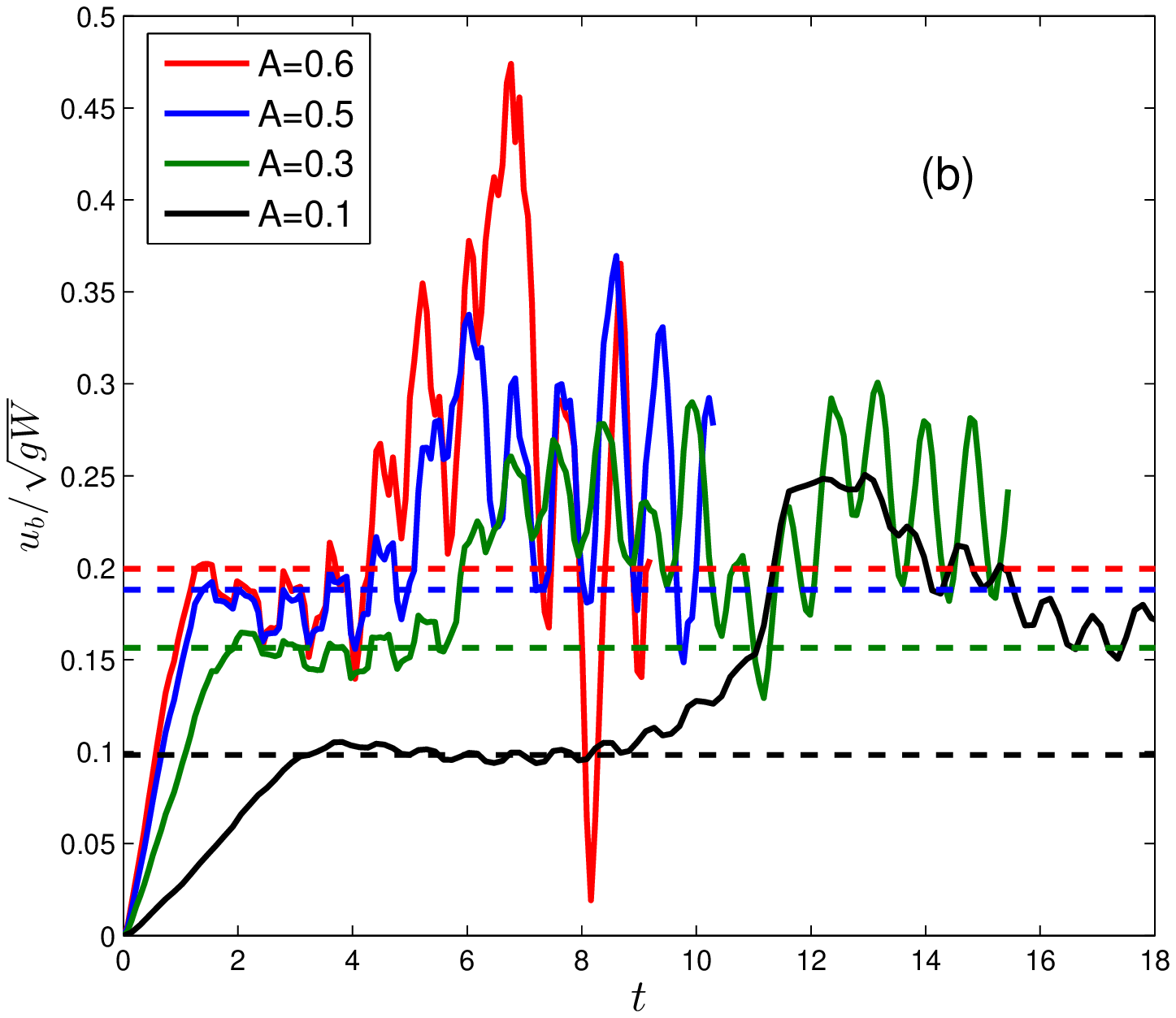}
\tiny\caption{Time variations of the (a) normalized bubble amplitude and (b) normalized bubble velocity in the immiscible RTI with various Atwood numbers. The dashed lines represent the analytical solutions of the classic potential flow model~\cite{Goncharov}. }
\end{figure}
	
We also measured the time variations of the spike and bubble normalized velocities with various Atwood numbers and plotted the results in Figs. 8(a) and 8(b). From these figures, the development of the single-mode instability in regardless of the Atwood number can also be divided into four distinguishing stages. In the linear stage, the perturbation develops in the form of an exponential function with the spike growth factor, which shows an increase with the Atwood number. The basic trend that the bubble velocity at the linear stage increases with the Atwood number was also observed, but the relative growth was slowly decreased. For the cases of $A=0.6$ and $A=0.5$, the developments of the bubble velocities are very similar at the linear stage, and the curves are even almost overlapped at the saturated velocity stage. This indicates that when the Atwood number is large enough, it no longer has a significant effect on the bubble velocity at the early stage. In the saturated velocity stage, the spike and bubble show to grow with nearly constant velocities and their values are found to increase with the Atwood number. For comparisons, the asymptotic velocities of the potential flow theory~\cite{Goncharov} depending on the Atwood number are also shown in Figs. 8(a) and 8(b). It can be seen that all the numerical predictions for the spike and bubble saturated velocities agree well with the analytical solutions in general. In addition, we can also find that durations of the saturated velocity stages for both spike and bubble decrease with the increase of the Atwood number and specially the sustaining time for the case of 0.6 is very transient. We can anticipate that the saturated velocity stage for the spike would disappear if continue to enlarge the Atwood number. This is because the vortices are generated faster at a higher Atwood number that accelerates the spike and bubble velocities exceeding than the asymptotic solutions such that the maintenance time of the quasi-steady state becomes shorter. Following the saturated velocity stage, the spike and bubble velocities begin to diverge from the ones obtained from potential flow theory and the growth of the stability enters into the reacceleration stage. The reacceleration stage for the spike emerges at earlier time than that of the bubble, which is more obvious with the increase of the Atwood number. The discrete vortices are first produced in the vicinity of the spike, and there is a certain distance between the generated vortices and the bubble tip, which thus leads to the time lag of the bubble entering the reacceleration stage than the spike. Lastly, the spike and bubble velocities at the chaotic development stage are unstable, presenting a state of some fluctuations with time, while the fluctuant range shows an increase with the Atwood number.

\begin{figure}
\includegraphics[width=3.4in,height=2.8in] {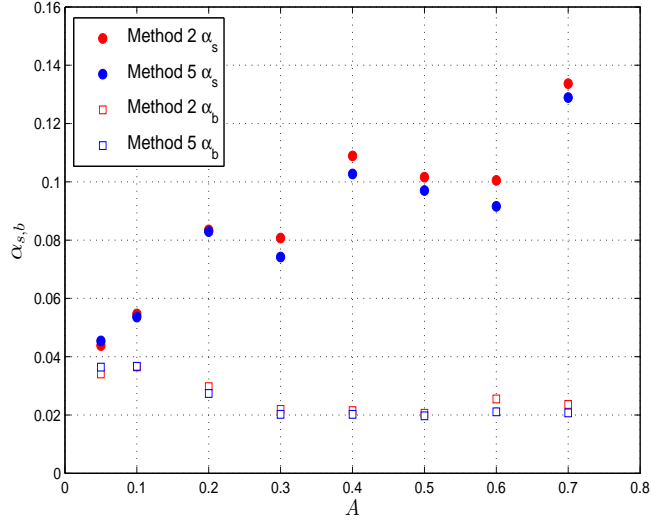}
\tiny\caption{Dependencies of the spike and bubble growth rates $\alpha_s$ and $\alpha_b$ on the Atwood number.}\label{fig:alpha}
\end{figure}
	
To reveal the late-time growth law of single-mode instability, we also measured the spike and bubble growth rates under a wide range of Atwood numbers from 0.05 to 0.7, which are the most concerned physical quantities in the study of interfacial instability phenomenon. We used two popular statistic methods to calculate the spike and bubble growth rates with different Atwood numbers and plotted the results in Fig. 9. It can be found that the computed growth rates obtained by the two methods are close to each other. For low Atwood number less than 0.1, the difference between the spike and bubble growth rates is small, which indicates the nearly symmetric developments of the spike and bubble fronts. However, for high Atwood numbers, the spike growth rate is remarkably larger than that of the bubble, implying the growth of the single-mode RTI becomes increasingly asymmetric about the positions of the spike and bubble interfaces. In addition, we can also observe from Fig. 9 that the spike growth rate shows an overall trend of the increase with the Atwood number, while the bubble growth rate has a slightly decrease with the Atwood number at first and then seems to be independence of the Atwood number approaching to a constant of 0.0215. It is worth noting that the predicted curves for spike and bubble growth rates of single-mode RTI
are very similar to the ones reported in the experimental studies of the multi-mode phenomenon~\cite{Akula, Dimonte, Banerjee1}.

\section{CONCLUSIONS}~\label{sec:summary}
In this paper, the late-time evolution of the immiscible single-mode RTI is studied by using the phase field lattice Boltzmann method, which mainly includes two aspects: (1) the influence of wide Reynolds numbers on the development of single-mode RTI with a moderate high Atwood number and (2) the influence of different Atwood numbers on the growth of single-mode RTI at a large Reynolds number. Numerical results show that the single-mode instability with a Atwood number of 0.7 presents distinct interfacial dynamics under different Reynolds numbers. For high Reynolds numbers, the phase interfaces roll up at multiple positions caused by the KH instability and eventually formed into a complicated topology structure in addition to numerous visible droplets. As the Reynolds number decreases, the interfaces become more and more smooth and even the roll-up or breakup phenomenon does not occur at extremely small Reynolds numbers. In addition, we summarized the development of the single-mode RTI with high Reynolds numbers into four different stages, including the linear growth, saturated velocity growth, reacceleration and chaotic development stages. The spike and bubble velocities at the saturated velocity stage are quantitatively consistent with the asymptotic solutions of the potential flow model, although the lasting time of the spike's stage is transient and it is expected that this stage for the spike would disappear by continuously enlarging the Atwood number. The spike and bubble amplitudes at the late-time chaotic stage exhibit quadratic growths characterized by two important parameters of the spike and bubble growth rates. Here we also conducted a comparison of several statistical approaches for computing the growth rate coefficients and two preferential methods are suggested. It is shown that the computed spike and bubble growth rates for a moderately high Atood number of 0.7 approach to 0.13 and 0.022, respectively. As the Reynolds number is reduced, some later stages cannot be reached sequentially. Four different growth stages can be also observed in the single-mode RTI with various Atwood numbers, while the time instants entering these stages are different, which is related to the propagation of vortex in the development process. Moreover, we calculated the growth rates of the spike and bubble under different Atwood numbers. For $A>0.2$, the bubble growth rate is basically maintained at a steady value of 0.0215, indicating that it is independence of the initial density ratio, while the spike coefficient shows an increase with the Atwood number in general. The present investigation only focused on the single-mode RTI between two incompressible fluids and the effect of compressibility on the late-time growth of RTI would be an interested subject~\cite{Lai, Luoxs, Wieland}, which constitutes one of our future works.

\section*{Acknowledgments}
This work is financially supported by the National Natural Science
Foundation of China (Grant No. 11972142), and the Natural Science Foundation of Zhejiang Province (No. LY19A020007).


\end{document}